\UseRawInputEncoding
\documentclass[fleqn,usenatbib]{mnras}
{\newif\ifnotend
\notendtrue
\def\veclist{ABCDEFGHIJKLMNOPQRSTUVWXYZabcdefghijklmnopqrstuvwxyz.}
\def\top#1#2.{#1}
\def\tail#1#2.{#2.}
\loop\expandafter\xdef\csname v\expandafter\top\veclist\endcsname%
{{\noexpand\bf\expandafter\top\veclist}}
\edef\veclist{\expandafter\tail\veclist}
\if\veclist.\notendfalse\fi\ifnotend\repeat}
{\newif\ifnotend
\notendtrue
\def\callist{ABCDEFGHIJKLMNOPQRSTUVWXYZ.}
\def\top#1#2.{#1}
\def\tail#1#2.{#2.}
\loop\expandafter\xdef\csname c\expandafter\top\callist\endcsname%
{{\noexpand\cal\expandafter\top\callist}}
\edef\callist{\expandafter\tail\callist}
\if\callist.\notendfalse\fi\ifnotend\repeat}

\def\Vc{v_{\rm c}}

\def\mnras{MNRAS}
\def\araa{AnnRA\&A}
\def\apj{ApJ}
\def\apjs{ApJS}
\def\apjl{ApJL}
\def\nat{Nat}

\def\aap{A\&A}

\def\bolth{\mbox{\boldmath$\theta$}}

\def\Gyr{\,\mathrm{Gyr}}

\def\kpc{\,\mathrm{kpc}}

\def\rg{R_{\mathrm{g}}}
\def\kms{\,\mathrm{km\,s}^{-1}}

\def\msun{\,{\rm M}_\odot}

\def\pc{\,\mathrm{pc}}

\usepackage{newtxtext,newtxmath}
\usepackage{amsmath,fleqn,graphicx,xcolor}
\usepackage{multirow}
\renewcommand{\[}{\begin{equation}}
\renewcommand{\]}{\end{equation}}
\def\i{{\rm i}}

\usepackage[T1]{fontenc}

\DeclareRobustCommand{\VAN}[3]{#2}
\let\VANthebibliography\thebibliography
\def\thebibliography{\DeclareRobustCommand{\VAN}[3]{##3}\VANthebibliography}

\usepackage{graphicx}	
\usepackage{amsmath}	

\title[Gaia phase spiral]{Gaia DR3 features of the phase spiral and its possible relation to internal perturbations}

\author[Li et al.]{
Chengdong Li,$^{1}$\thanks{E-mail: chengdong.li@astro.unistra.fr}
Arnaud Siebert,$^{1}$
Giacomo Monari,$^{1}$
Benoit Famaey,$^{1}$
Simon Rozier,$^{2}$
\\
$^{1}$Universit\'e de Strasbourg, CNRS, Observatoire astronomique de Strasbourg (ObAS), UMR 7550, 67000 Strasbourg, France\\
$^{2}$School of Mathematics, University of Edinburgh, Kings Buildings, Edinburgh, EH9 3FD, UK \\
}
\date{Accepted XXX. Received YYY; in original form ZZZ}

\pubyear{}

\begin{document}
\label{firstpage}
\pagerange{\pageref{firstpage}--\pageref{lastpage}}
\maketitle

\begin{abstract}
Disc stars from the Gaia DR3 RVS catalogue are selected to explore the phase spiral as a function of position in the Galaxy. The data reveal a two-armed phase spiral pattern in the local $z-v_z$ plane inside the solar radius, which appears clearly when colour-coded by $\langle v_R \rangle (z,v_z)$: this is characteristic of a breathing mode that can in principle be produced by in-plane non-axisymmetric perturbations. We note the phase spiral pattern becomes single armed outside the solar radius. When a realistic analytic model with an axisymmetric background potential plus a steadily rotating bar and 2-armed spiral arms as perturbation is used to perform particle test integrations, the pseudo stars get a prominent spiral pattern in the $\langle v_R \rangle$ map in the $x-y$ plane. Additionally, clear breathing mode evidence at a few $\kms$ level can be seen in the $\langle v_z \rangle$ map on the $x-z$ plane, confirming that such breathing modes are non-negligible in the joint presence of a bar and spiral arms. However, no phase-spiral is perceptible in the $(z, v_z)$ plane. When an initial vertical perturbation is added to all pseudo stars to carry out the simulation, the one-armed phase spirals can clearly be seen 500~Myr after the perturbation and gradually disappear inside-out. Finally, we show as a proof of concept how a toy model of a time-varying non-axisymmetric in-plane perturbation with varying amplitude and pattern speed can produce a strong two-armed phase-spiral. We conclude a time-varying strong internal perturbation together with an external vertical perturbation could perhaps explain the transition between the two-armed and one-armed phase-spirals around the Solar radius.


\end{abstract}




\section{Introduction}{\label{sec:intro}}

The phase spiral structure in the vertical position and velocity ($z,v_z$) plane of the Milky Way is one of the most important discoveries since the launch of the Gaia mission \citep{GaiaCollaboration2016}. This feature was first revealed in the solar vicinity with a population of stars sharing similar Galacto-centric radii and azimuthal angles \citep{Antoja2018}. In that work, the spiral is prominent in the $(z,v_z)$ plane colour-coded by $\langle v_\phi \rangle$, the mean value of the streaming velocity in the direction of Galactic rotation, but not very clear in stellar density $\rho$, or when colour-coded by $\langle v_R \rangle$, the mean radial velocity in Galactic cylindrical coordinates. Some subsequent works, however, showed that spirals can also be clearly seen in the stellar density \citep{Laporte2019,Li2021,GaiaCollaboration2022a} and colour-coded by 
$\langle v_R \rangle$ \citep{Binney2018} or angular momentum $\langle L_z \rangle$ \citep{Khanna2019}.

Since then, different attempts were carried out to elucidate the origin and longevity of this spiral feature. Usually, the phase spiral is supposed to be generated by an external perturber such as a dwarf galaxy merging with the Milky Way.  \citet{Binney2018}, for instance, developed a toy model based on the impulse approximation to show that the perturbation imposed by an intruder (such as a dwarf galaxy or some dark matter subhalo) can cause a prominent phase spiral in $\langle v_\phi \rangle(z,v_z)$.  \citet{Darling2019} used both analytical methods and numerical simulations to model the pure bending wave in the disc, obtaining that a passing satellite will cause the disc to bend and will generate spirals in the $(z,v_z)$ plane coloured by median $v_R$ or $v_\phi$. Similarly, \cite{Laporte2019} analysed an $N$-body simulation of the interaction between the Milky Way and a Sagittarius-like dwarf spheroidal galaxy, showing that it can produce a phase spiral, including in the $\rho(z,v_z)$ map. \citet{BlandHawthorn2021} also carried out a high-resolution $N$-body simulation of an impulsive mass interacting with a cold stellar disc for a single transit, showing that the phase spiral was excited by a dwarf galaxy transiting the disc 1-2 Gyr ago. Another two simulations were carried out by \citet{Hunt2021} to study the merger and evolution history of the Milky Way. Both one-armed and two-armed phase spirals are witnessed in these simulations, which indicates that both bending and breathing modes can be triggered by the interaction with an external perturber. 

A concurrent explanation of the phase spiral associates it with internal mechanisms within the disc. In particular, \citet{Khoperskov2019} used an $N$-body simulation of an isolated Milky Way-type galaxy to show that the vertical oscillations driven by bar buckling can generate phase spirals in the solar neighbourhood. 
Additionally, \citet{Grand2022} carried out a cosmological magnetohydrodynamic simulation to investigate the formation of disequilibrium disc structures, concluding that the phase spiral is induced by the dark matter wakes which lead to the formation of the disc warp. While this is not an internal disc mechanism, it nevertheless does not rely on a single external impact from the perturber but is building up over time through the response of the dark matter halo. In a similar vein, \citet{Tremaine2022} showed that the phase spiral could result from Gaussian noise in the gravitational potential, potentially related to to substructures in the dark-matter halo.

Recently, using Gaia DR3 observations of inner disk stars, \citet{Hunt2022} spotted a two-armed spiral feature in $\rho(z, v_z)$ when stars are binned by guiding radius $\rg$ and $\theta_\phi$, the conjugate angle of the action $J_\phi$ (the vertical component of the angular momentum). Interestingly, they also found such a two-armed phase spiral in an isolated simulation. Meanwhile, a two-armed phase spiral was also measured in $\langle v_R \rangle(z,v_z)$ inside the solar radius by \citet{Antoja2022}. In a theoretical study based on linear perturbation theory, \citet{Banik2022} showed that such two-armed features --- related to breathing modes --- are prominently produced by relatively fast, more impulsive perturbations. Recently, the framework of the shearing
box approximation was adopted by \cite{Widrow2023} to investigate the response of a disc to an impulsive excitation in the presence of self-gravity of the disc and swing amplification. They concluded that a two-armed phase spiral pattern can be triggered by an excitation that is symmetric about the mid plane.

The phase space can also be reformulated from the $z-v_z$ plane into the frequency-angle $\Omega_z-\theta_z$ plane, which transforms the phase spirals into a `zebra diagram' \citep[e.g.]{Li+2021}. Here, $\Omega_z$ is the orbital frequency of vertical motion and $\theta_z$ is the conjugate angle to the vertical action $J_z$. The stripes in the $\Omega_z-\theta_z$ plane can be related to specific perturbation ages \citep[e.g.,][]{Li+2021}. Recently, \cite{Frankel2022} developed a method to fit the straight lines in the $\Omega_z-\theta_z$ plane, whose slope has the dimension of inverse time, to time the perturbations: they found a perturbation age varying between 0.2 and 0.6 Gyr. \cite{Widmark2022} previously found consistent times (between $\sim$ 0.35 and 0.767 Gyr) with a similar method, but using stars in different volumes in the Galaxy. In that case, the phase spiral was fitted directly in the $(z,v_z)$ plane, adjusting both the time and the parameters of the background axisymmetric potential at the same time. Finally, \cite{Tremaine2022} proposed that those stripes could also be generated by many small disturbances rather than a large one.  

In light of the recently discovered two-armed phase spiral inside the solar radius, which can be produced by internal plane-symmetric perturbations, this paper aims at analysing what kind of phase spiral could potentially be produced by the bar and spiral arms of the Galaxy. To that end, we refine the representations of the phase spirals along the disc on a finer grid, to try to characterize the transition region from one-armed to two-armed phase spirals within the disc. We then compare the data, for the time being at the {\it qualitative} level, to test-particle simulations in a  galactic potential with different bar and spiral arms perturbations added on top of a self-consistent background axisymmetric model.

The paper is organized as follows. In Section~\ref{sec:obs}, we describe the features of the two-armed spiral pattern in phase space when colour-coded by $\langle v_R \rangle(z,v_z)$ with the Gaia DR3 RVS sample. Then, our models for the axisymmetric galaxy and its perturbations are introduced in Section~\ref{sec:theo}: we exhibit the perturbations produced by a steadily rotating bar and a steadily rotating density wave for the spiral arms, and we qualitatively compare them with the observations in Section~\ref{sec:simcom}. In Section~\ref{sec:per}, we add an initial vertical perturbation to the test particles and demonstrate the response of those particles to the perturbation by the bar and spiral arms. We finally show in Section~\ref{sec:str} how a toy model of non-axisymmetric in-plane perturbations with varying amplitude and pattern speed can in principle produce a strong two-armed phase-spiral. Section~\ref{sec:cons} summarizes and discusses directions for future plans.

\section{Phase space spirals with Gaia DR3}{\label{sec:obs}}

\subsection{The Gaia DR3 RVS sample}{\label{sec:data}}
The stellar sample in this work is selected from the third release of the Gaia catalogue \citep{GaiaCollaboration2022b}. The typical uncertainties for parallax and proper motion are 0.02-0.03 mas and 0.02-0.03 mas yr$^{-1}$ at $G = 9 \sim 14$, 0.07~mas and 0.07~mas~yr$^{-1}$ at $G = 17$~mag, and 0.5~mas and 0.5~mas yr$^{-1}$ at $G = 20$~mag for 5-parameter solutions \citep{Lindegren2021}. In addition, Gaia DR3 includes more than 33 million stars with radial velocity measurements, which have $G_{\text{RVS}}$ $\leq$ 14 and with 3100 $\leq$ $T_{\text{eff}}$ $\leq$ 14500 K \citep{GaiaCollaboration2022b}. 

We select the stars with $\varpi/\epsilon_{\varpi}\,>\,5$, $G > 5$, and RUWE $< 1.4$. The magnitude cut here is to avoid the saturation of the brightest end of Gaia observations. The RUWE is the Renormalised Unit Weight Error, which is a scale of the quality of the astrometric solution. The stars' distances are from the $distance\_gspphot$ from the DR3 catalogue, which uses the GSP-Phot Aeneas best fit values \citep{Andrae2022}. 

\begin{figure}
	\includegraphics[width=\columnwidth]{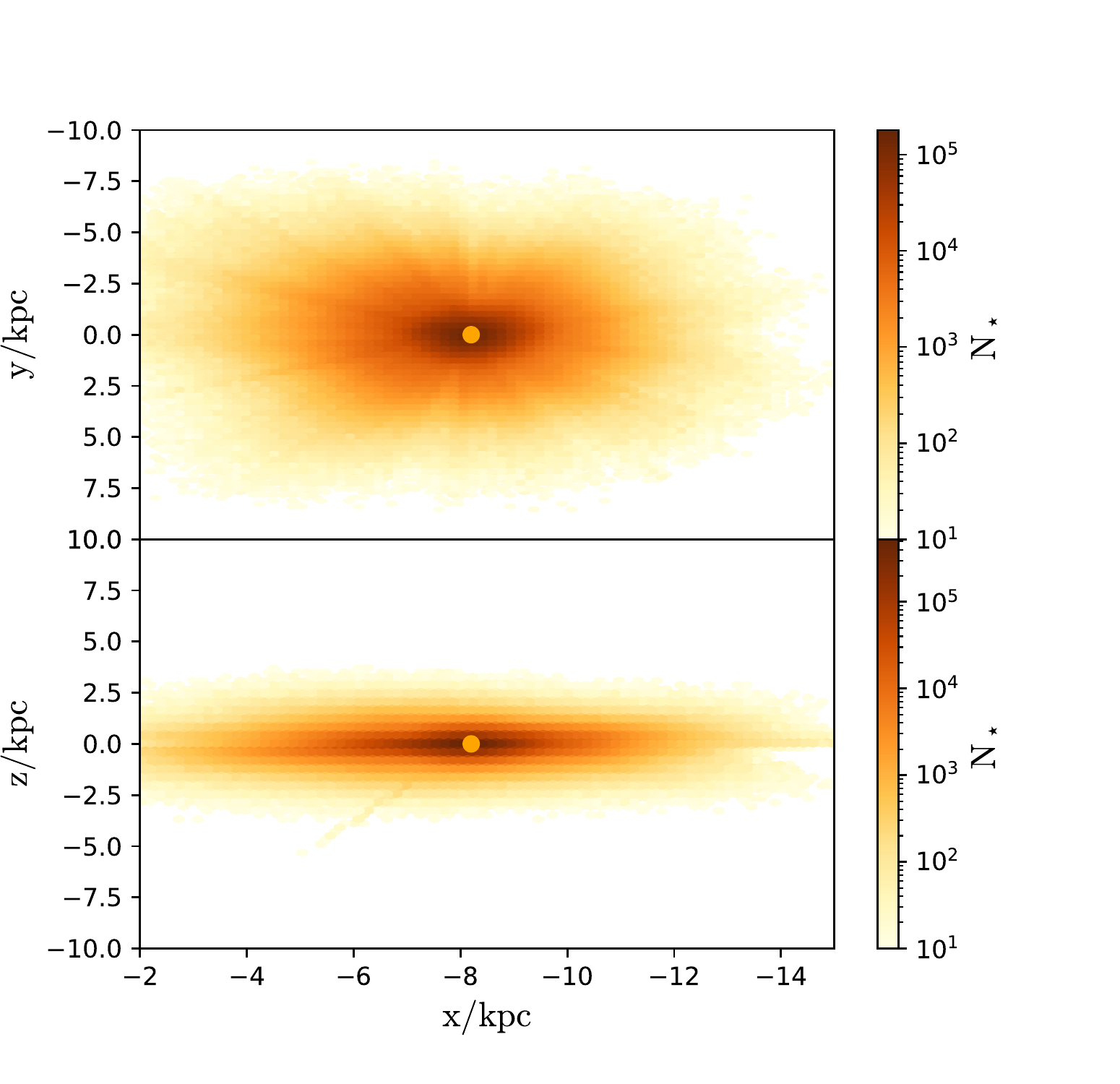}
	\vspace{-0.3cm}
    \caption{The spatial distribution of selected stars in the Gaia catalogue in the $x-y$ and $x-z$ plane respectively. $\mbox{N}_\star$ denotes the number of stars in each bin in the plots.}
    \label{fig:star_dis}
    \vspace{-0.3cm}
\end{figure}

Figure~\ref{fig:star_dis} shows the stellar density distribution of our sample in the $x-y$ and $x-z$ planes respectively. The orange dot denotes the Sun with coordinates $(x, y, z) = (-8.2, 0, 0.02)~\kpc$. The positions and velocities in Cartesian and cylindrical coordinates are computed using $Astropy$ \citep{Astropy2022}. The Sun's velocity with respect to the LSR is taken to be $(U_\odot,V_\odot,W_\odot)$ = (11.1, 12.24, 7.25) $\kms$ \citep{Schonrich2010} and the LSR velocity is $V_{\text{LSR}}$ = 238 $\kms$ \citep{Schonrich2012}. 

The actions $\textbf{J}$ and conjugate angles $\bolth$ are computed from $(\textbf{x},\textbf{v})$ via the `St\"{a}ckel Fudge', which is based on assuming that the Galaxy's potential is a  St\"{a}ckel potential \citep{Binney2012}. The $AGAMA$ package \citep{Vasiliev2019} is adopted to compute the actions and angles. The potential used to derive the actions is generated from a self-consistent Galaxy model introduced in detail in \citet{Binney2023}.

\subsection{Bending mode or breathing mode?}{\label{sec:inter}}

\begin{figure*}
	\includegraphics[scale=0.45]{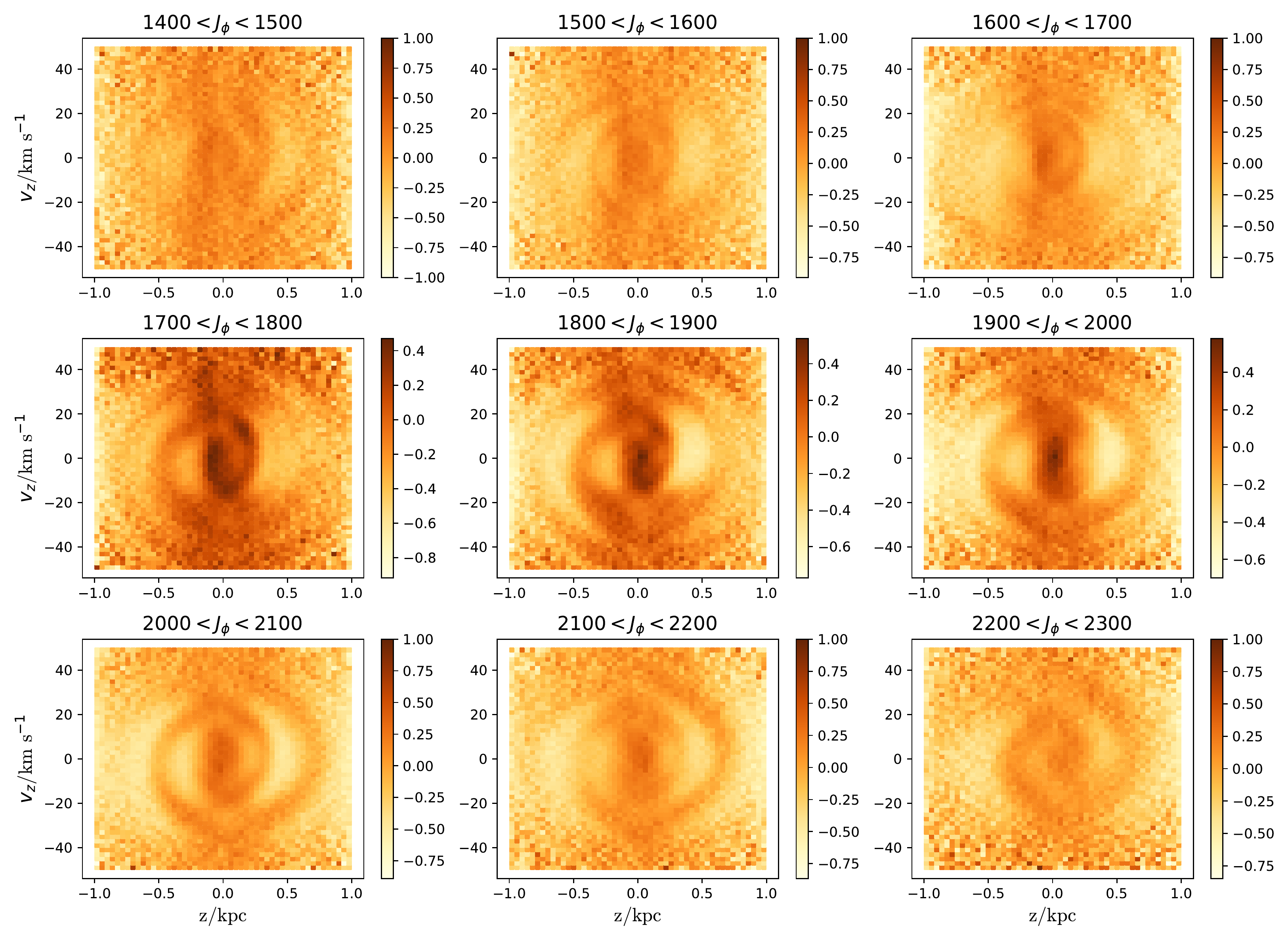}
	\vspace{-0.3cm}
    \caption{The phase space density plot for our selected Gaia RVS sample data. The plotted quantity is the relative overdensity w.r.t. an average local density, following a procedure introduced in the text in detail. Only one armed phase spirals are seen in different $J_\phi$ intervals.}
    \label{fig:phase_dens}
    \vspace{-0.3cm}
\end{figure*}

Figure~\ref{fig:phase_dens} shows the density plot of the sample stars in the $z-v_z$ phase space binned by the value of their angular momentum (azimuthal action) $J_{\phi}$\footnote{The absolute values of $J_{\phi}$ are shown on top of the figures in this work, as we use a right-handed coordinate system and the angular momentum of stars with prograde orbits is negative.}. The overdensity map is shown, using the same method as in \citet{Laporte2019}: stars are binned in the $(z, v_z)$ space in pixels of size $\Delta \mbox{pix}=0.04\times2\,(\mbox{kpc}\times\mbox{km}\,\mbox{s}^{-1})$. The resulting density $\rho(z, v_z)$ is then smoothed by convolution with a Gaussian filter of width $4\,\Delta \mbox{pix}$, yielding the smoothed density $\bar{\rho}(z, v_z)$. We then plot the overdensity $\hat{\rho}=\rho(z, v_z)/\bar{\rho}(z, v_z)-1$.

Different panels correspond to different azimuthal action bins, which is an increasing function of the guiding radius $\rg$: $J_\phi=\rg \times \Vc(\rg)$. As a consequence, the 9 panels in Figure~\ref{fig:phase_dens} are approximately sorted from the inner to the outer Galaxy. It should be noticed that the plots of the first row with the lowest angular momenta bins only include stars within 1.5 kpc from the Sun, which aims at eliminating the influence of dust extinction towards the Galactic centre.  

A single spiral pattern appears clearly in most panels, although this is not clear for the three innermost (lowest angular momenta) bins due to the large scatters of the data points in these bins. This one-armed phase spiral, as discussed thoroughly in the literature, could be related to a bending mode in the local and outer Galaxy, which could be triggered by the impact of the Sagittarius dwarf galaxy \citep[e.g.,][]{Binney2018,Laporte2019,BlandHawthorn2021}. 

\begin{figure*}
	\includegraphics[scale=0.45]{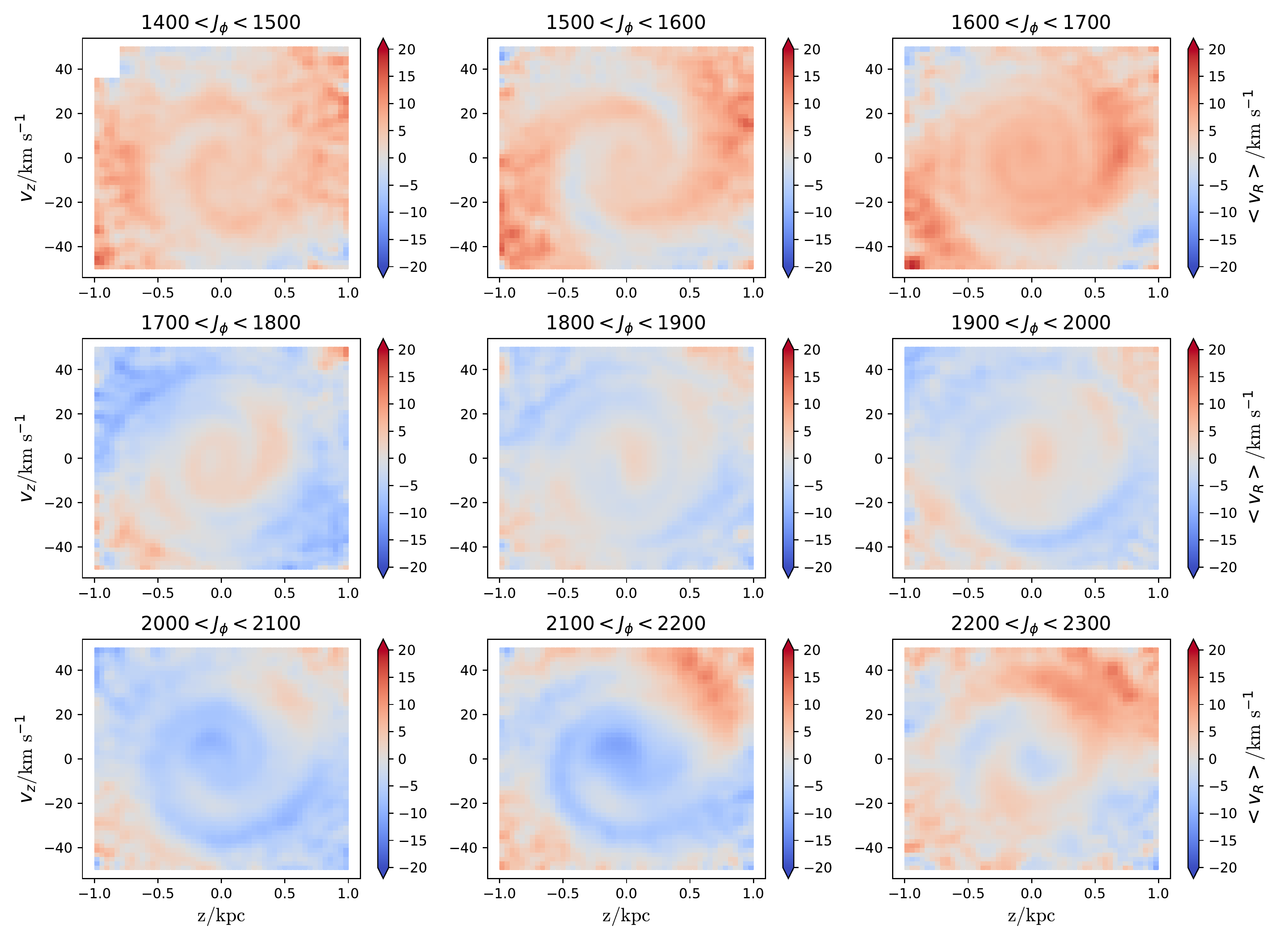}
	\vspace{-0.3cm}
    \caption{The phase space plot colour-coded by $\langle v_{R}\rangle$ for our selected Gaia RVS sample data. This plot is also Gaussian smoothed with the same width $4\,\Delta \mbox{pix}$ for both $z$ and $v_z$ axes. Unlike Figure~\ref{fig:phase_dens}, what we plot here is $\overline{v_R}(z, v_z)$. Two armed phase spirals are clearly seen inside the solar radius when $J_\phi\,<\,1800$. The phase spiral pattern becomes single armed when $J_\phi\,>\,1800$.}
    \label{fig:phase_vr}
    \vspace{-0.3cm}
\end{figure*}

Several previous works have shown that the phase spiral can sometimes appear more clearly, and with different characteristics, either in the $\langle v_R \rangle(z, v_z)$ and $\langle v_\phi \rangle(z, v_z)$ maps \citep[e.g. ][]{Antoja2018,Binney2018,Laporte2019,BlandHawthorn2021}. Figure~\ref{fig:phase_vr} therefore displays the mean Galactocentric radial velocity maps $\langle v_R \rangle(z,v_z)$ in different $J_{\phi}$ bins. The results are particularly interesting in the inner Galaxy. The first four panels, with $1400 \, \text{km~s}^{-1}\text{~kpc}< J_\phi < 1800 \, \text{km~s}^{-1}\text{~kpc}$, indeed show  clear {\it two-armed} spiral patterns. When $J_\phi>1800$ km~s$^{-1}$~kpc (next five panels), only one-armed spiral features can be seen in the plots, as in the density maps of Fig.~\ref{fig:phase_dens}. The two-armed spirals in $\langle v_R \rangle(z,v_z)$ at low $J_{\phi}$, also seen in Figure 4 of \citet{Antoja2022} around $\rg\,\simeq\,6.7\,\kpc$, showing that the inner Galaxy probably experiences a different perturbation history from that in the outer galaxy. The two-armed phase-spiral pattern corresponds to a breathing wave \citep[e.g.,][]{Widrow2023} and is a strong hint for another perturbation mechanism than the perturbation causing the one-armed spiral in the outer disc \citep{Hunt2022}. 

Although sample stars are selected from the same Gaia DR3 RVS catalogue in both this work and that of \citet{Hunt2022}, the two-armed spiral at small guiding radius in the density plot discovered in their work is not seen here. One possible reason is that they not only bin the stars by guiding radius $\rg$ but also by angles $\theta_\phi$, which can emphasize the spiral features along specific azimuthal angles, while we marginalized over azimuth. Secondly, the detailed selection criteria are different between this work and that of \citet{Hunt2022}, which can also lead to disparities among the $z-v_z$ phase space plots. On the other hand, the phase space features shown in \citet{Antoja2022} are similar to those in this work, including the one-armed spirals in the density plots and two-armed spirals in some specific bins of the $\langle v_R \rangle(z, v_z)$ map. The difference is that the bin size in \citet{Antoja2022} is twice as large as that in this work, which reduces the number of bins that can detect two-armed spirals in the $\langle v_R \rangle(z, v_z)$ map.

\begin{figure*}
	\includegraphics[scale=0.45]{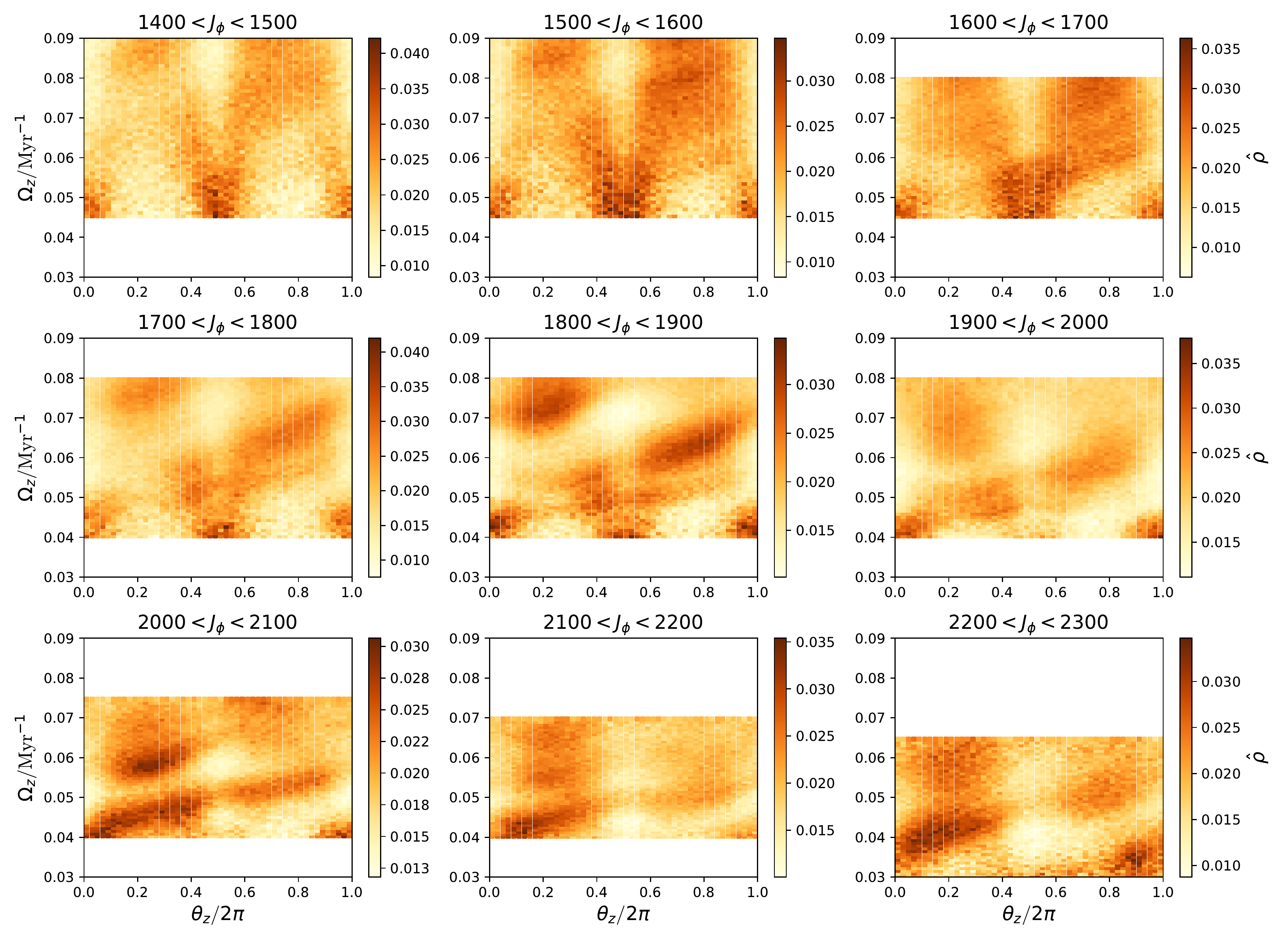}
	\vspace{-0.3cm}
    \caption{The density distribution of the stars in the $\Omega_z-\theta_z$ plane. The re-normalized density is $\hat{\rho}_{ij}=\mbox{N}_{ij}(\theta_z,\Omega_z)/{\sum_{j}\,\mbox{N}_{ij}(\theta_z,\Omega_z})$. Some stripes can be seen on each panel with positive slopes. The slopes of each stripe has a relation with the perturbation time scale.}
    \label{fig:thetazomegaz}
    \vspace{-0.3cm}
\end{figure*}

Figure~\ref{fig:thetazomegaz} shows the density distribution of the stars in the $\Omega_z-\theta_z$ space. The re-normalized density for a given pixel is the number count of points in this pixel divided by the total number of points at this frequency $\Omega_z$, thus $\hat{\rho}_{ij}=\mbox{N}_{ij}(\theta_z,\Omega_z)/{\sum_{j}\,\mbox{N}_{ij}(\theta_z,\Omega_z})$, where $\mbox{N}_{ij}(\theta_z,\Omega_z)$ is the number count in a given pixel in the $\Omega_z-\theta_z$ space. Some clear "zebra" stripes are observed on each panel of this plot. For the plots in the first row, the stripes are in the shape of two arches. From the fourth panel, the shape changes to one or two straight lines with positive slopes. These stripes shape can in principle be interpreted as the consequence of a perturbation to the distribution function at time $t=0$, with the shape as $\Delta f(0)=A\,\cos{[m(\theta_z-\theta_{z0})]}$. After a time $t$, the perturbation becomes $\Delta f(t)=A\,\cos{[m(\theta_z-\Omega_z t-\theta_{z0})]}$, where $\theta_{z0}$ is the phase angle, $m$ represents the index of the Fourier component, and $\Omega_z t$ shows the change of the conjugate angle $\theta_z$ since the event. The extreme values of this function are lines with slope $1/t$ on the $\Omega_z-\theta_z$ plane \citep{Frankel2022,Tremaine2022,DarraghFord2023}. On the contrary, the arch-like stripes at low $J_\phi$ probably result from selection effects due to dust obscuration \citep{Frankel2022}. 

As a conclusion, while the density map shows only one-armed spiral patterns in the outer Galactic disc, which corresponds to a bending mode, no clear two-armed spiral can be detected in the inner Galaxy. However, the $\langle v_R\rangle(z,v_z)$ map shows clear two-armed spiral patterns in the inner Galaxy with guiding radii between about $\sim$6.0 to $\sim$7.7 kpc. 

This result raises the following questions: is it possible to have multiple triggers, both internal and external, to stimulate the phase spirals, and do the stars respond differently in $\rho$ and $v_R$? In order to investigate the possible relationship between the observed two-armed phase spiral features in the $v_R$ map and the internal perturbations, we will now model the bar and spiral arms of the Galaxy and analyse the response of the test particles to the perturbations in Section~\ref{sec:theo}. 

\section{Modelling predictions under perturbations}{\label{sec:theo}}

\subsection{Axisymmetric Galaxy model and test particle sampling}{\label{sec:model}}

In order to help interpret the observations presented in the previous section, we will now model the response of the Galactic disc to internal plane-symmetric perturbations using a particle-test approach. 
The Galaxy's model used in this work contains two parts, the axisymmetric background and the non-axisymmetric perturbation. 
The idea of the particle-test approach is generating the test particles based on the axisymmetric distribution functions (DFs), then integrating the pseudo stars' orbits in a potential including the non-axisymmetric perturbations. On the one hand, the advantage of this approach is that it is adaptable to adjust the models producing the perturbations and it takes only a few hours to run a simulation with 10 million test particles. Any response of the test particles to the perturbations can be detected easily in both densities and kinematics based on each snapshot. On the other hand, the limitation of this approach is that self-gravity is ignored, while it has been shown that it could play a role in shaping or sustaining phase spirals \citep[e.g.][]{Khoperskov2019,Widrow2023}.
In addition, we do not simulate the selection function of the Gaia RVS sample precisely. The results of the simulation can be only considered to mimic the real situation in the Galaxy approximately. Our present goal is therefore not to quantitatively fit the observations, but to obtain qualitative insights on the effect of internal non-axisymmetries on phase-spirals based on this test particle approach.

The DF based axisymmetric Galaxy model that we construct with the $AGAMA$ code \citep{Vasiliev2019} consists of a dark halo, a stellar halo, a bulge and a four-component stellar disc. 
The disc component consists in the young, middle-age and old thin discs, and a high-$\alpha$ disc. The DF of each component is a specified function $f(\bf{J})$ of the action integrals. In addition, a gas disc which is not included in the DF model is added to derive the total potential of the Galaxy. The details of the instruction of this Milky Way self-consistent model and the various predictions by this model can be found in \citet{Binney2023}. 

\begin{figure}

	\includegraphics[width=\columnwidth]{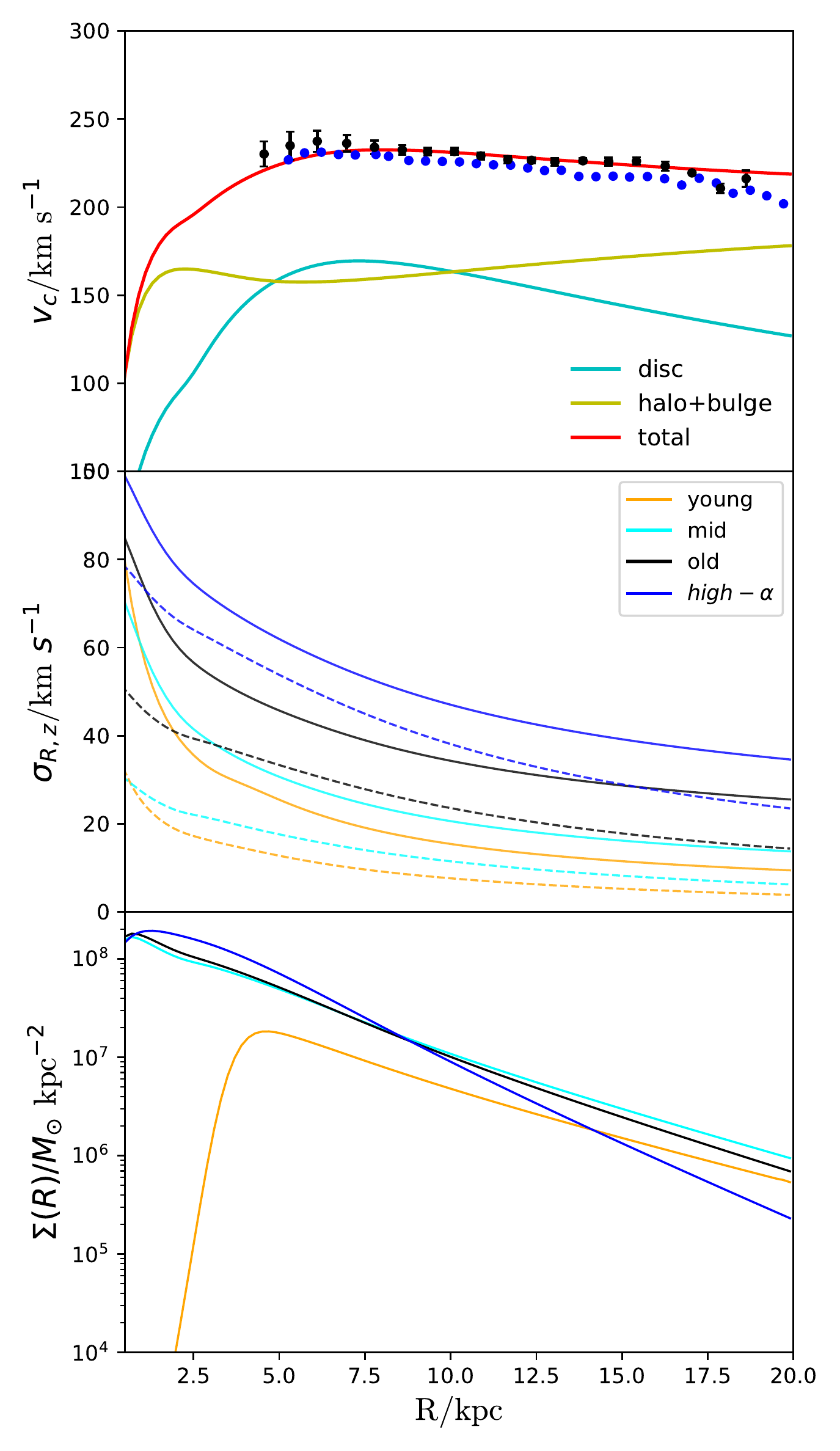}
    \vspace{-0.6cm}    
    \caption{The circular velocity within 20 $\kpc$ from the Galactic centre is shown in the upper panel, in which the yellow curve represents the sum of all the spheroidal components of the model including the bulge and halo. The red curve denotes all four disc components. The blue and black dots are from \citet{Eilers2019} and \citet{Ablimit2020} respectively. The middle and lower panels respectively show the velocity dispersion and surface density profiles of the different disc components from which the pseudo stars are sampled. In the middle panel, from top to bottom the velocity dispersion curves represent the high-$\alpha$ disc, the old thin disc, the middle-age thin disc and the young thin disc. The solid and dashed curves represent the velocity dispersion profiles in $\sigma_R$ and $\sigma_z$ respectively. The surface density of the young thin disc, which is in orange in the lower panel, shows a declining trend towards both the centre and the outskirts, as we applied a tapering of the DF at a distance of about 4 kpc from the Galactic centre.}
    \label{fig:dens_disp}
    \vspace{-0.7cm}
\end{figure}

The model used in this work is however slightly different from the model in \citet{Binney2023}. For the dark halo, stellar halo and the bulge, we use a simple double power-law DF model which is the same as \citet{Li2022a}. Meanwhile, for the young thin disc, a taper-exponential model is adopted according to \citet{Li2022b}. The circular velocity of this axisymmetric potential is displayed in in the first panel of Figure~\ref{fig:dens_disp}. According to such a model, the disc components have a maximum circular velocity at $6-8 \kpc$ away from the centre.    

Finally, 10 million disc stars are sampled based on the DFs, which includes 2 million young thin disc stars and high-$\alpha$ stars, 3 million middle age thin disc stars and old age thin disc stars respectively. This choice of pseudo star numbers is due to the relative fraction of each disc components. The normalization of the thick disc, which we use as the high-$\alpha$ component in this work, at $(R,z)=(R_\odot, 0)$ is 0.10-0.15 \citep{Juric2008}. We choose 0.2 as the fraction for the high-$\alpha$ disc in this work since our test particles represent the whole Galaxy \citep{BlandHawthorn2016}. The ratios of the three thin disc components are roughly based on the normalization parameters in the self-consistent model. The DF used in this work and the Galaxy model can be found in the online supplementary material. The velocity dispersions and surface densities of all four disc components are displayed in the middle and lower panels of Figure~\ref{fig:dens_disp}. In the middle panel, the velocity dispersions for various disc components are shown with the solid lines denoting $\sigma_{R}$ and dashed lines denoting $\sigma_{z}$. In the lower panel, the surface density curves all show exponential profiles at large radii.

\subsection{Steadily rotating bar and spiral arm model}{\label{sec:bar}}

The internal perturbation that we first consider for the Galaxy includes two parts, a central bar and spiral arms. 

The bar, rotating at a given pattern speed $\Omega_{\rm{b}}$, is modelled following \citet{Chiba2022} as
\begin{equation}
    \Phi_{\rm{b}}(r,\theta,\phi,t)\,=\,\Phi_{\rm{br}}(r)\sin^2{\theta}\cos{m(\phi-\Omega_{\rm{b}} t)},
    \label{eq:phib}
\end{equation}
where $(r,\theta,\phi)$ are the spherical coordinates and we consider $m\,=\,2$ for the bar, i.e. we include only the quadrupole term. The radial dependence of the bar potential $\Phi_{\rm{br}}(r)$ is
\begin{equation}
    \Phi_{\rm{br}}(r)\,=\,-\frac{A\,\Vc^{2}}{2}\,\bigg(\frac{r}{r_{\rm{CR}}}\bigg)^{2}\,\bigg(\frac{b+1}{b+r/r_{\rm{CR}}}\bigg)^{5},
    \label{eq:phibr}
\end{equation}
where $A$ quantifies the strength of the bar, $\Vc$ is the value of circular velocity in the solar vicinity, i.e. $\Vc\,=\,235\,\kms$. The parameter $b$ describes the ratio between the bar scale length and the value of the co-rotation radius $r_{\rm{CR}}$. The parameters used in this work are given by \citet{Chiba2022}, with $A\,=\,0.02$ and $b\,=\,0.28$. The parameters are all summarized in the upper panel of Table~\ref{tab:bar_spipp}.
\begin{table}
	\centering
	\vspace{0.4cm}
	\caption{The parameters used for the bar and spiral arms in the model with steady rotation. $\Omega$ is in $\kms\,\kpc^{-1}$, and $\Vc$ in $\kms$. $r_{\rm{CR}}$, $R_{\rm{s}}$ and $h_{\rm{s}}$ are in $\kpc$. $\phi_{\rm{b}}$ and $\phi_0$ denote the initial phase angles of the bar and the spiral arms respectively. The unit of $\Sigma_0$ is $\msun\,\kpc^{-2}$.}
	\label{tab:bar_spipp}
	\begin{tabular}{lccccccc} 
		\hline
		Bar  & $\Omega_{\rm{b}}$ & $A$ & $\Vc$ & $b$ & $r_{\rm{CR}}$  & $\phi_{\rm{b}}$\\
	
		Values & -35 & 0.02 & 235 & 0.28 & 6.7 & $172^{\circ}$\\
            \hline
		Spiral arm &  $\Omega_{\rm{p}}$ & $R_{\rm{s}}$ & $h_{\rm{s}}$ & $N$ & $\alpha$ & $\phi_0$  &  $\Sigma_0$\\
		
		Values &   -18.9 &1.0 & 0.1 & 2 & $9.9^{\circ}$ & $103^{\circ}$ & $2.5\times10^9$\\
		\hline
	\end{tabular}
	\vspace{-0.6cm}
\end{table}

The spiral arm model used in this work is the one from \citet{Cox2002},

\begin{equation}
    \Phi_{\rm{s}}(R,\phi,z)\!=\!-4\pi G \Sigma_0\mathrm{e}^{-R/R_{\rm{s}}}\!\!\sum_n\!\frac{C_n}{K_n\,D_n}\!
    \cos n\gamma \big[\!\cosh\!\big(\tfrac{K_nz}{\beta_n}\big)\!\big]^{-\beta_n}\!\!,
    \label{eq:spiralarm}
\end{equation}
where $(R,\phi,z)$ are cylindrical coordinates, and $\Sigma_0$ is the central surface density. $C_{n=1,2,3}$ are ${C_1=8/3\pi,C_2=1/2,C_3=8/15\pi}$ representing the amplitudes of the three harmonic terms and the functional parameters are
\begin{equation}
    \begin{aligned}
    &K_n\,=\,\frac{nN}{R\sin{\alpha}},\\
    &\beta_n\,=\,K_n\,h_{\rm{s}}(1+0.4K_n h_{\rm{s}}),\\
    &\gamma\,=\,N\bigg[\phi\,-\,\frac{\ln{(R/R_{\rm{s}})}}{\tan{\alpha}}\,-\,\Omega_{\rm{p}} t\,-\,\phi_0 \bigg],\\
    &D_n\,=\, \frac{1}{1+0.3K_nh_{\rm{s}}}\,+\,K_nh_{\rm{s}},
    \end{aligned}
    \label{eq:spiral_paras}
\end{equation}
where $N$ is the number of arms, $h_{\rm{s}}$ is the scale height, $\alpha$ is the pitch angle, $\phi_0$ is the phase, and $n=1,2,3$ denotes the three harmonic terms. The parameters used in the spiral arm potential can be found in the lower panel of Table~\ref{tab:bar_spipp}, most of which are from \citet{monari2016a,monari2016b} denoting a tightly wound spiral pattern. However, the pattern speed and phase angle have the opposite sign of \citet{monari2016a,monari2016b} as we use a right-handed coordinate system in this work. The surface density we choose keeps the spiral arm potential with a radial force of about 0.5 per cent compared with the axisymmetric background potential in the solar vicinity \citep{Siebert2012}. The example of this perturbation potential file used in this work can be found in online supplementary files.

The pattern speed of the bar is $\Omega_{\rm{b}}\,=\,-35\,\kms\,\kpc^{-1}$ \citep{Binney2020,Chiba2021b} in this coordinate system\footnote{We use a right-handed coordinate system in this work, and since the bar rotates in the same sense as disk stars, its pattern speed is also negative. The pattern speed of spiral arms is also the same. In this coordinate system, the disc, bar, and spiral arms all rotate clockwise.}, and the pattern speed of the $N=2$ spiral arms is set to be $\Omega_{\rm{p}}\,=\,-18.9\,\kms\,\kpc^{-1}$ \citep{monari2016b} without changing throughout the orbit integration. The total Galaxy's potential file can be found in the online supplementary material. 

We start the simulation by evolving the pseudo stars in the axisymmetric background potential for 1 $\Gyr$ without any perturbation. Then we add the bar and spiral arms potential at $T = 1.0 \Gyr$, and do the orbital integration for another 4 $\Gyr$. In summary, the orbital integration routine in $AGAMA$ is adopted to compute the trajectories of all the mock stars over 5 $\Gyr$ and the trajectories are stored every 0.02 $\Gyr$. We simulate the phase angles of the bar and spiral arms so that they approximate their current orientation w.r.t the Sun at the end of the simulation: at $T\,=\,5\,\Gyr$, the bar is assumed to have a phase angle $\phi\,=\,28^{\circ}$ \citep{Wegg2015} and the spiral arms are assumed to have a phase angle $\phi\,=\,26^{\circ}$ \citep{monari2016b}. The initial values of the phase angles of the bar and spiral arms are presented in Table~\ref{tab:bar_spipp}.

\subsection{Results for steadily rotating perturbations}{\label{sec:simcom}}

\begin{figure}
    
	\includegraphics[width=\columnwidth]{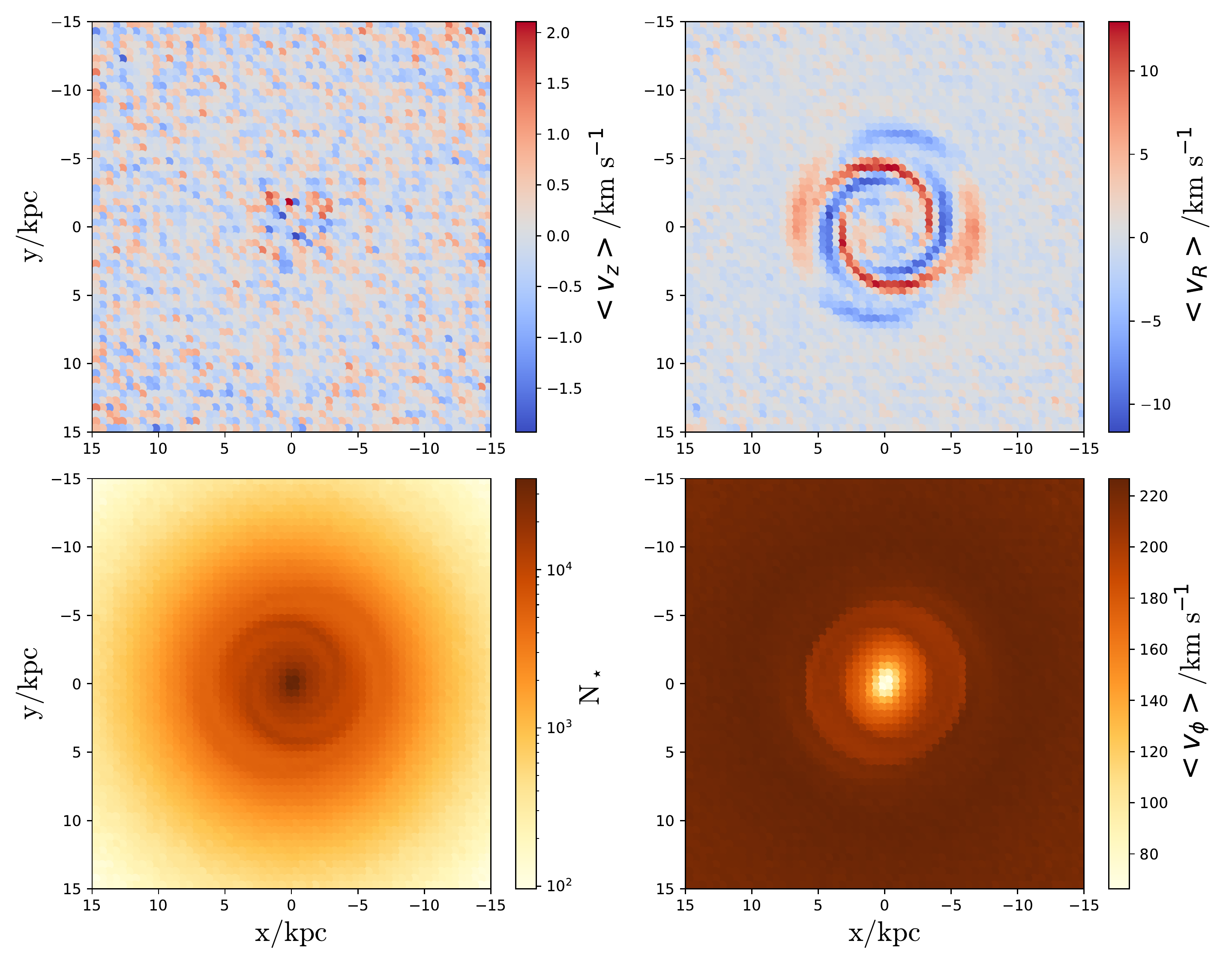}
    \vspace{-0.3cm}    
    \caption{The distribution of the density $\rho$, $\langle v_R \rangle$, $\langle v_\phi \rangle$, and $\langle v_z \rangle$ in the $x-y$ plane at $T\,=\,5.0\Gyr$ of the steadily rotating perturbers model. The perturbation in the $x-y$ plane is clear to see in $\langle v_R \rangle$, $\langle v_\phi \rangle$ and $\rho$. The mean vertical velocity in the $x-y$ plane is nearly zero everywhere due to symmetry w.r.t. the $z=0$ plane.}
    \label{fig:car_05gyr}
    \vspace{-0.3cm}
\end{figure}

The stellar distribution and velocity patterns in Cartesian coordinates within the disc plane are shown in Figure~\ref{fig:car_05gyr} for mock stars at $T = 5.0 \Gyr$. The density plot shows an elliptical central dense region and two-armed spiral features outwards. The spiral features on $\langle v_R \rangle$ are more prominent. The shape of the spiral arms is visible across the disc, and this perturbation on $\langle v_R \rangle$ is as predicted by \citet{monari2016a, monari2016b}, both in terms of amplitudes and location with respect to the the under- and over-densities.

\begin{figure}
	\includegraphics[width=\columnwidth]{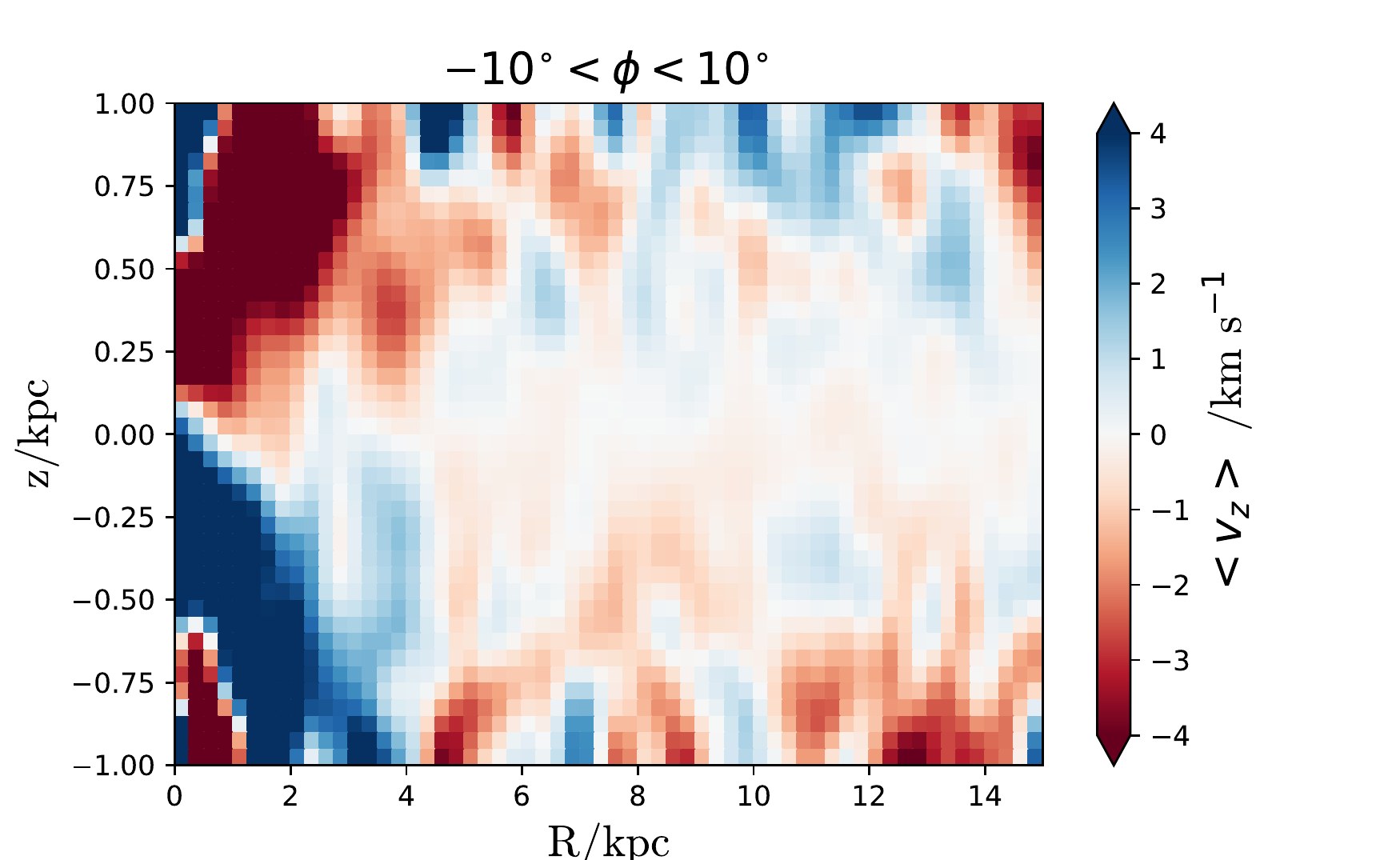}
	\vspace{-0.3cm}
    \caption{The distribution of $\langle v_z \rangle(R, z)$ for pseudo stars at $\rm{T}\,=\,2.0\,\Gyr$ at azimuthal angles $-10^\circ<\phi<10^\circ$. The result of a breathing mode triggered by the bar and spiral arms is visible for the stars with $R\,<\,5\kpc$ and $R\,>\,10\kpc$. This figure is smoothed with a Gaussian filter of $4\,\Delta \rm{pix}$, with $\Delta \mbox{pix}=0.25\times0.05\,(\kpc\times\kpc)$. The azimuthal angle $\phi=0$ coincides with the $Y=0$ axis from the Galactic centre to the opposite direction of the solar position. }
    \label{fig:xzvz}
    \vspace{-0.3cm}
\end{figure}

\begin{figure}
	\includegraphics[width=\columnwidth]{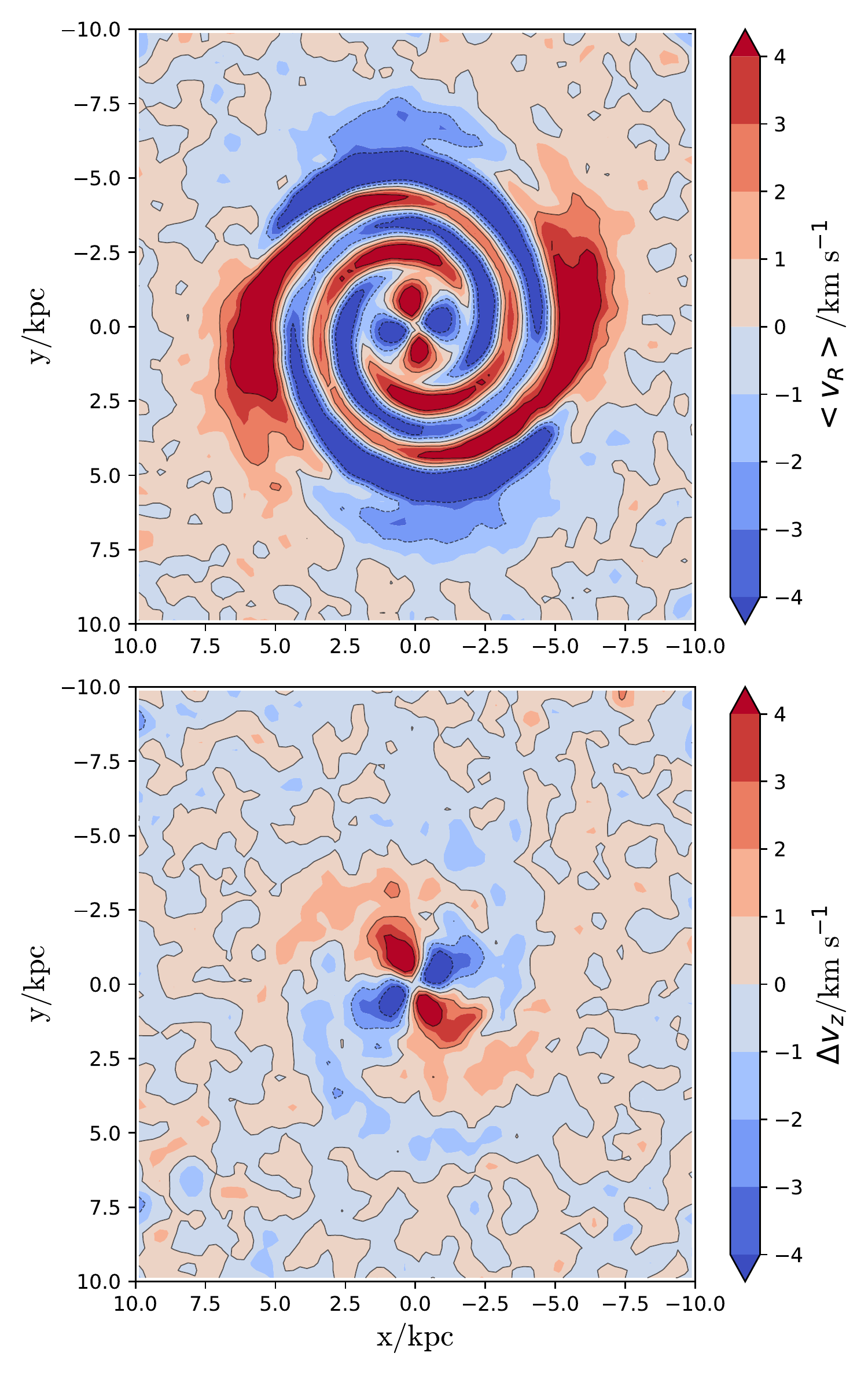}
	\vspace{-0.3cm}
    \caption{The face-on contour map of $\langle v_R \rangle$ and $\Delta v_z$ for the test particles at $\rm{T}\,=\,2.0\,\Gyr$ in the simulation with steadily rotating non-axisymmetries. The  value of $\Delta v_z$ is the difference between $\langle v_z \rangle$ for $z > 0$ and for $z < 0$ respectively. This map is Gaussian smoothed with a filter width of $4\,\Delta\mbox{pix}$, with $\Delta \mbox{pix}=0.25\times0.25\,(\kpc\times\kpc)$.}
    \label{fig:contour}
    \vspace{-0.3cm}
\end{figure}

What we plot as $\langle v_z \rangle$ in this plot is the mean value of vertical velocities including stars both above and below the plane: it is consistent with zero within Poisson noise, since the bar and spiral arms are plane-symmetric and only excite breathing modes and not bending modes. The projected mean vertical velocity is thus averaged to 0 when summing in $z$. In Figure~\ref{fig:xzvz}, the mean velocity in $v_z$ is shown in the meridional plane above and below the plane for pseudo stars with azimuthal angle $\phi$ between $-10^{\circ}$ and $10^{\circ}$. The velocity pattern as a consequence of the breathing mode perturbation can be seen in this figure above and below the Galactic plane for $R\,<\,5\,\kpc$ and $R\,>\,10\kpc$. The face-on contour maps of $\langle v_R \rangle$ and $\Delta v_z$ for the test particles at $\rm{T}\,=\,2.0\,\Gyr$ are shown in Figure~\ref{fig:contour}. The value of $\Delta v_z$ is the difference between $\langle v_z \rangle$ for $z > 0$ and for $z < 0$ respectively. The quadrupole mode of the bar can be seen explicitly in the breathing mode close to the centre of the Galaxy. Two tightly wound spiral arms are also clear to see. The quadrupole mode of the bar is constrained within $3 \kpc$ from the centre because we use a rather slow bar pattern speed $\Omega_{\rm{b}}\,=\,-35\,\kms\,\kpc^{-1}$. The face-on density map shows that the central bar and spiral arms can produce a breathing mode perturbation to the disc stars \citep{Faure2014,monari2015,monari2016a,monari2016b}.

\begin{figure*}
	\includegraphics[scale=0.45]{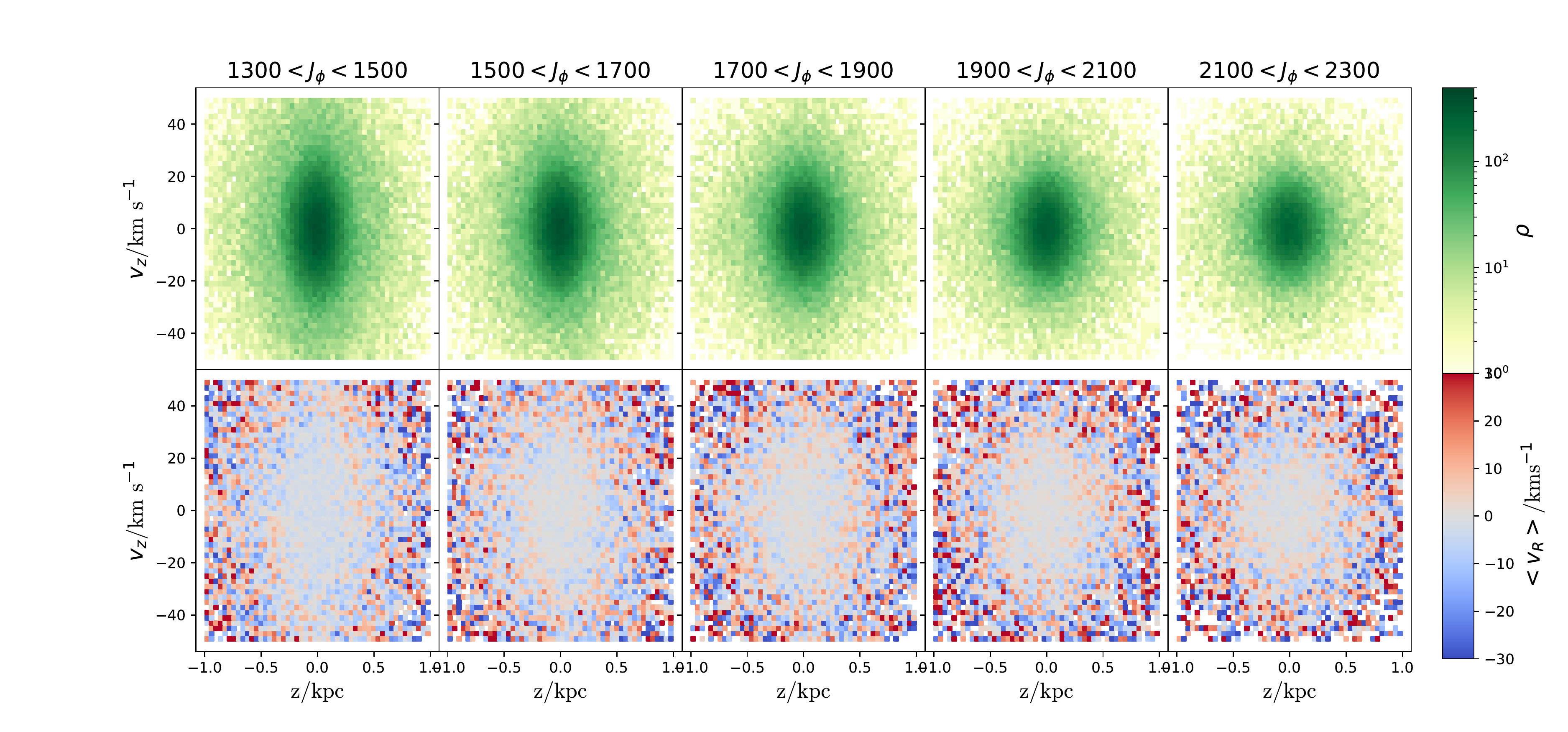}
	\vspace{-0.3cm}
    \caption{Maps of $\rho(z, v_z)$ (top) and 
$\langle v_R \rangle(z, v_z)$ (bottom) for the model including a bar and spiral arms with constant pattern speeds, at $T\,=\,1.5\,\Gyr$. The density $\rho(z, v_z)$ shows no spiral arm features. $\langle v_R \rangle(z, v_z)$ does not show any signals of the phase spiral either. Note that here, and more generally in all figures from simulations, no Gaussian smoothing is applied. 
Densities represent number counts of stars in phase space bins.}
    \label{fig:phase_05gyr}
    \vspace{-0.3cm}
\end{figure*}

Figure~\ref{fig:phase_05gyr} shows the phase space maps of $\rho(z, v_z)$ (top row) and
$\langle v_R \rangle(z, v_z)$ (bottom row) for different positions along the disc at $T\,=\,1.5\,\Gyr$. Figure~\ref{fig:phase_05gyr} (and the subsequent Figure~\ref{fig:thetazomegaz_mock}) only contains pseudo stars within $6\kpc$ from the Sun, which aims at making it roughly compatible with Gaia's observational region, assuming that the Sun is located at $(x,y,z)\,=\,(-8.2,0,0.02)\kpc$. 
Unlike the observational data, the density plot $\rho(z, v_z)$ shows no spiral arm features among different azimuthal action intervals. Each plot has a vertically elongated elliptical density distribution centered at around $(z,v_z)\,=\,(0,0)$. The vertically elongated shape is constrained by the initial conditions when sampling the pseudo stars. At $T=1.5\,\Gyr$, the potential of the central bar and spiral arms does not produce a phase spiral pattern in the density plot among the pseudo stars on the $z-v_z$ plane in this simulation. Furthermore, the $\langle v_R \rangle(z, v_z)$ plots do not show any sign of phase spirals either. The distribution of $\langle v_R \rangle(z, v_z)$ is smooth within the central region of the $z-v_z$ plane and gets progressively noisy outwards. We did test the pseudo stars phase space properties for other snapshots at various time steps in the simulation: there are no phase spiral structures either in the $(z, v_z)$ plane, whether it is in density or colour-coded by velocities. 

\begin{figure}
	\includegraphics[width=\columnwidth]{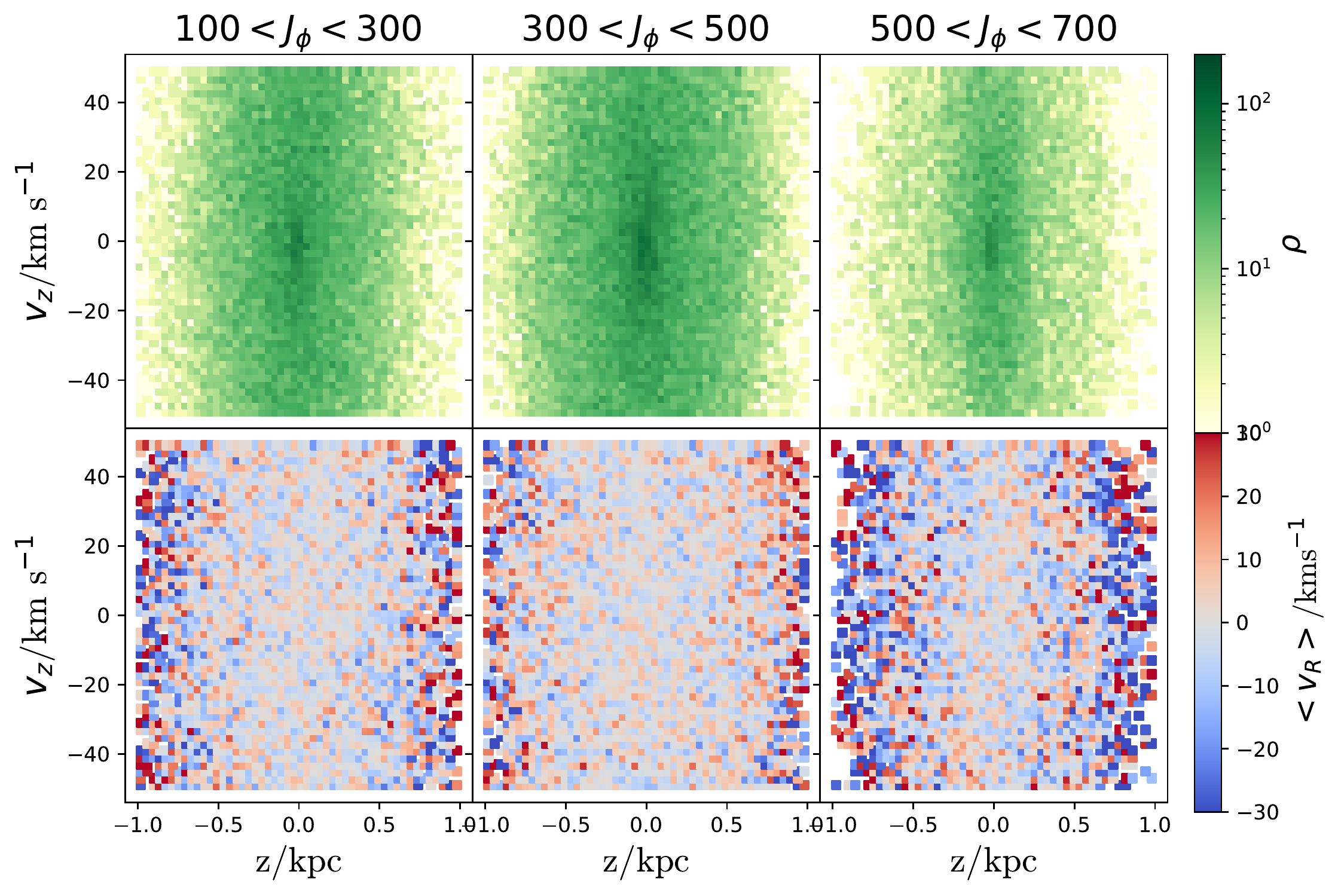}
	\vspace{-0.3cm}
    \caption{Maps of $\rho(z, v_z)$ (top) and $\langle v_R \rangle(z, v_z)$ (bottom) for pseudo stars selected in a square with $-2.5<x<0\,\kpc$ and $0<y<2.5\,\kpc$ at $T=1.5$~Gyr for the model including a bar and spiral arms with constant pattern speeds,. There are still no phase spirals in this plot. However, the ellipsoids in the centre are slightly inclined among all the panels, which is a sign of a breathing mode produced by the bar with constant pattern speed.}
    \label{fig:phase_inner}
    \vspace{-0.3cm}
\end{figure}

\begin{figure*}
	\includegraphics[scale=0.45]{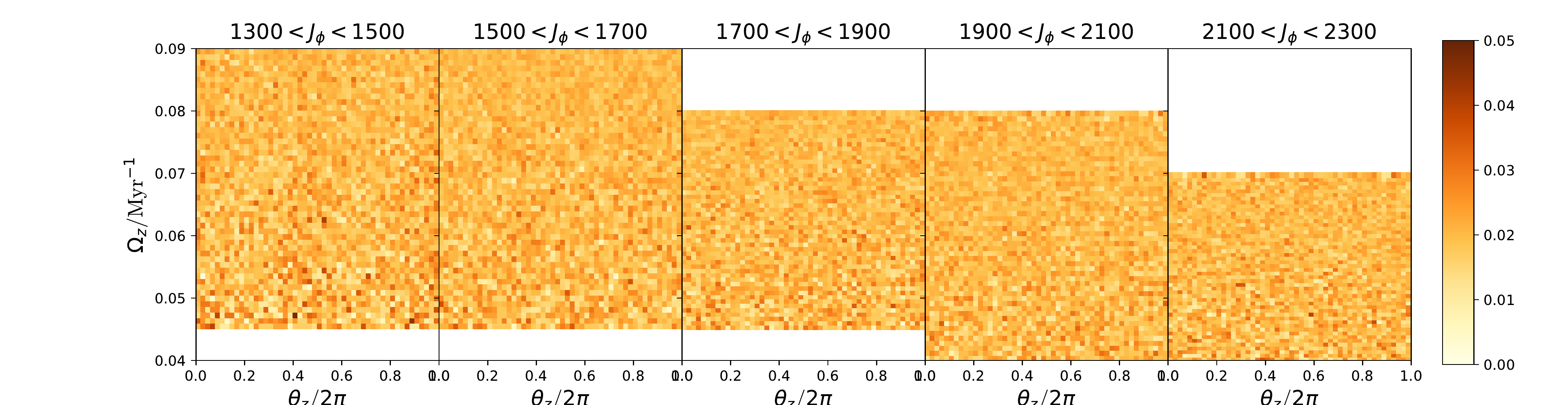}
	\vspace{-0.3cm}
    \caption{Pseudo-star densities in the $\theta_z-\Omega_z$ plane for the model where the bar and spiral arms have constant pattern speeds, at $T\,=\,1.5\,\Gyr$. The re-normalization method is the same as in Figure~\ref{fig:thetazomegaz}. No obvious stripes can be observed in these diagrams.}
    \label{fig:thetazomegaz_mock}
    \vspace{-0.3cm}
\end{figure*}

In order to explore the breathing mode shown in Figure~\ref{fig:car_05gyr} and Figure~\ref{fig:contour}, the density plot of $\rho(z, v_z)$ and $\langle v_R \rangle(z, v_z)$ for the pseudo stars at $T=1.5$~Gyr in the {\it inner disc}, where the breathing mode is most visible, is shown on Figure~\ref{fig:phase_inner}. The breathing mode corresponds to a compression-rarefaction wave and the zone where pseudo-stars were selected in the $x-y$ plane (see caption) has been chosen such as to be dominated by rarefaction (expansion). If we follow this zone in the frame rotating with the bar at later time-steps ($T=3.0$~Gyr, $T=4.0$~Gyr), it appears that the amplitude of the expansion $\Delta v_z$ does not evolve with time, indicating that pseudo-stars have reached near-equilibrium in the bar's rotating frame. It is clear from the plot in  $\rho(z, v_z)$ that the ellipsoid acquires an inclination, clearest in the third panel. This is a sign of the breathing mode produced by the perturbations with constant pattern speed, but, at near-equilibrium within the bar's steadily rotating potential, there is no phase mixing happening and no phase spiral in the $z-v_z$ plane. 

Figure~\ref{fig:thetazomegaz_mock} shows the density distribution of the stars in the $\Omega_z-\theta_z$ plane at $T\,=\,1.5\,\Gyr$ with the same re-normalization method as in Figure~\ref{fig:thetazomegaz}. As a reminder, the observational plots display prominent straight lines with positive slopes, which should have a strong correlation with the timing of the perturbation on the specific disc regions. On the contrary, no clear stripe patterns can be seen for the test particles in this simulation. Again, it means that, despite the presence of a breathing mode, the pseudo stars are at equilibrium within the potential of the background and steadily rotating perturbers, with no clear signs of phase mixing.

As a summary, according to Figure~\ref{fig:xzvz} and Figure~\ref{fig:contour}, the breathing mode produced by the bar and spiral arms is clear to see in both the $\langle v_z \rangle(R,z)$ and face-on contour maps. Nonetheless, only an inclination of the ellipsoid in the $z-v_z$ plane is observed and no phase spiral structures can be seen. In addition, the stripe patterns in $\Omega_z-\theta_z$ space cannot be seen in the pseudo stars either.

\subsection{Initial vertically perturbed test particles}{\label{sec:per}}

In Section~\ref{sec:simcom}, we showed that the perturbation produced by a central bar and spiral arms with a constant pattern speed cannot stimulate two-armed phase spiral structures among pseudo stars sampled from an axisymmetric Galactic distribution. It is therefore worthwhile to investigate how the pseudo stars evolve in this potential when an initial vertical perturbation is imposed, mimicking an interaction with an external perturber. Can the joint endeavors of this external perturbation and of the internal non-axisymmetric perturbations produce a two-armed phase spiral in the central parts and a one-armed phase spiral in the outer parts? 

For this, we displace every pseudo star by a random $\delta_z$, sampled from a Gaussian distribution centred on $\delta_z=400\,\pc$ with dispersion $\sigma_{\delta_z}=150\,\pc$. We also kick the initial velocities by a random $\delta_{v_z}$, sampled from a Gaussian distribution centred on $5 \kms$ with  dispersion $2\kms$. Thus the original positions and velocities in the $z$-direction for test particles all have some shifts away from the axisymmetric background, which is similar to that used in \citet{Antoja2018}. In the present toy-model, we apply this kick to the whole disk, although only the region around the Sun will be extracted: we analyze the test particles within $6\kpc$ from the Sun in order to mimic Gaia observations. The orbital integration scheme is the same as in the previous simulation without the vertical kick.

\begin{figure*}
	\includegraphics[scale=0.45]{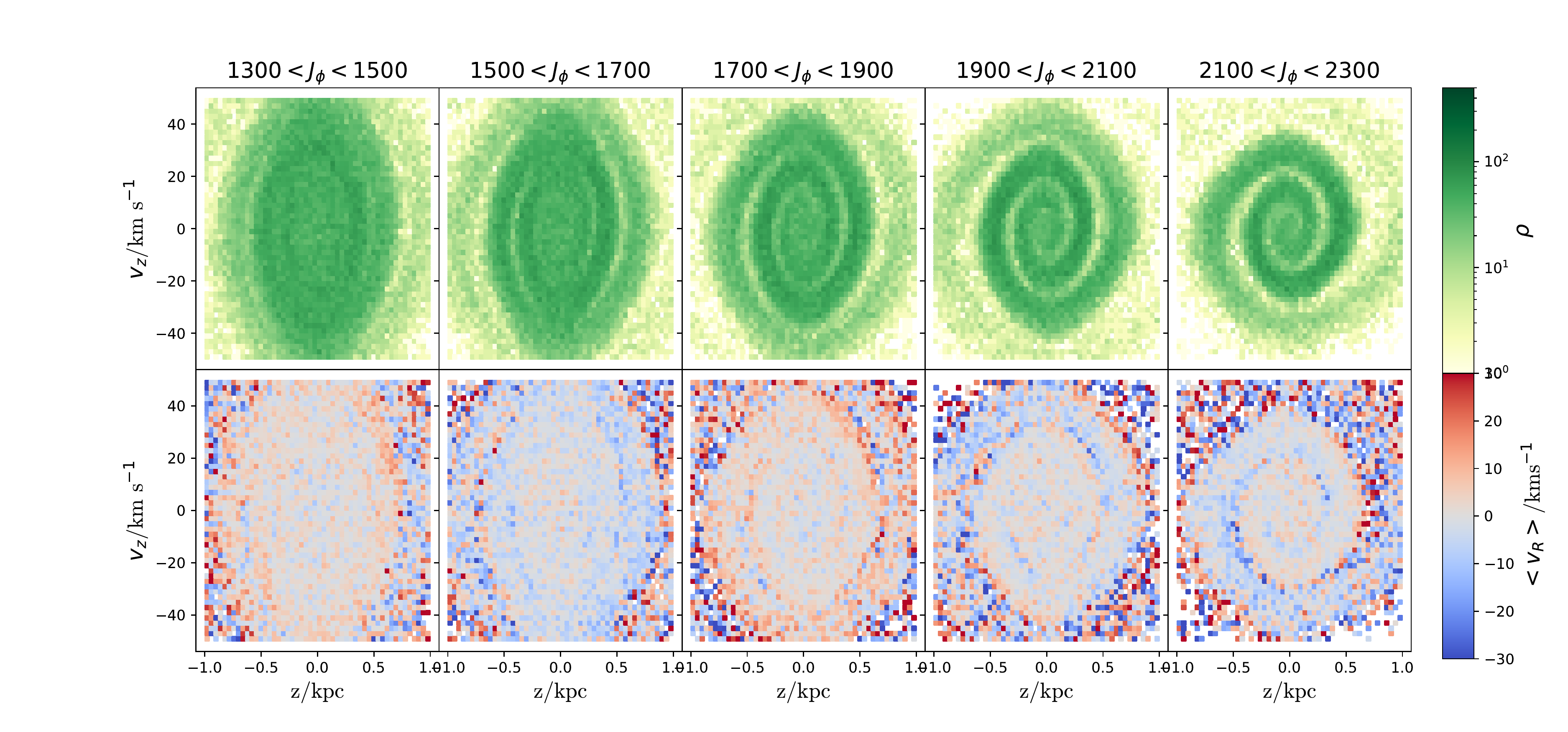}\\
	\vspace{-0.8cm}
        \includegraphics[scale=0.45]{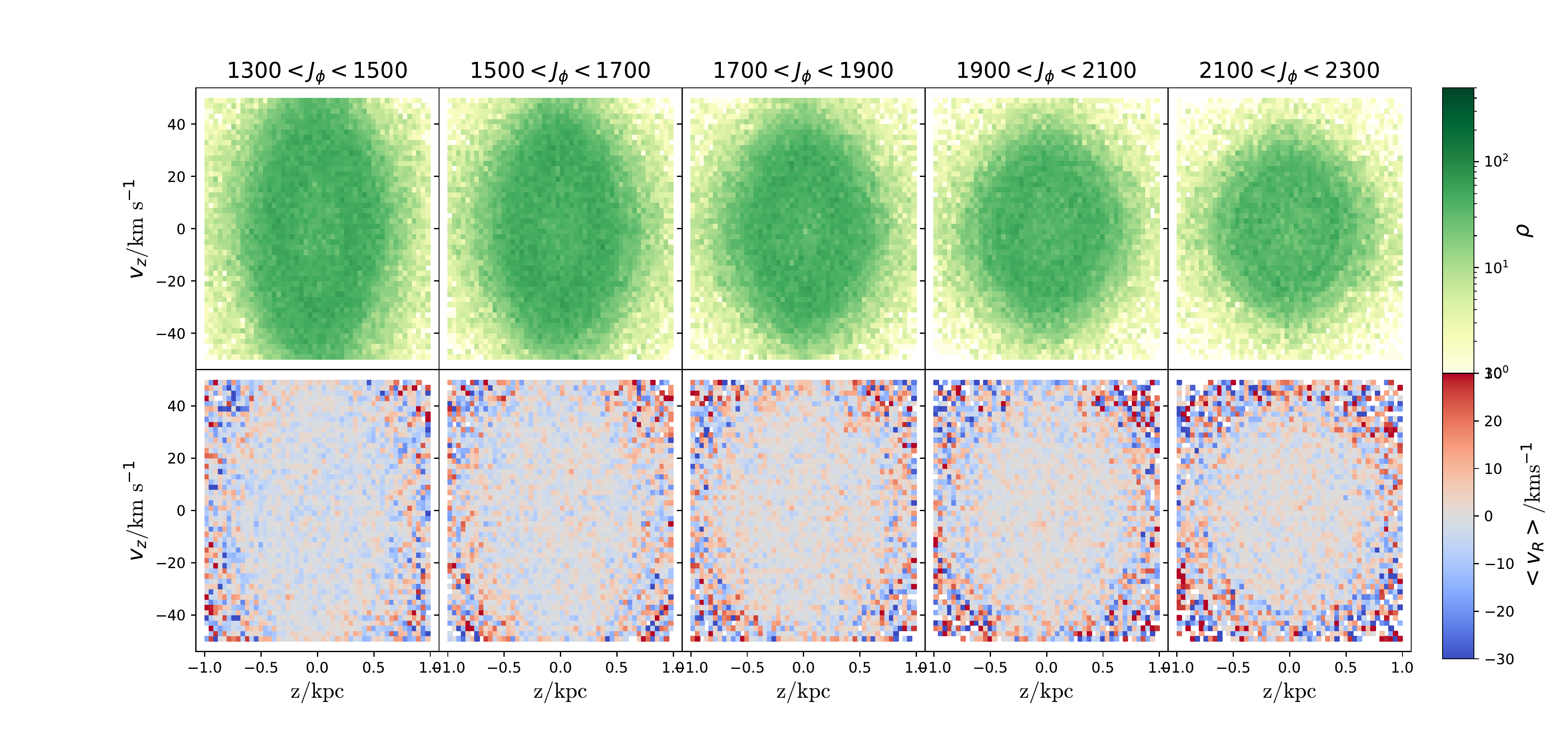}
    \vspace{-0.5cm}
    \caption{Maps of $\rho(z, v_z)$ and 
$\langle v_R \rangle(z, v_z)$ for test particles with an initially perturbed distribution, at $T\,=\,0.5\,\Gyr$ (upper grid) and $T\,=\,1.5\,\Gyr$ (lower grid). In each grid, the top row shows the density $\rho(z, v_z)$ and the bottom row $\langle v_R \rangle(z, v_z)$. At $T\,=\,0.5\,\Gyr$, both show single-armed phase spiral features. At $T\,=\,1.5\,\Gyr$, only the last panel of the density plot $\rho(z, v_z)$ shows spiral arm features now. The spiral structures in $\langle v_R \rangle(z, v_z)$ disappear.}
    \label{fig:phase_05gyr_pert}
    \vspace{-0.3cm}
\end{figure*}

The resulting phase space plots $\rho(z, v_z)$ and
$\langle v_R \rangle(z, v_z)$ at $T\,=\,0.5\,\Gyr$ are shown in Figure~\ref{fig:phase_05gyr_pert}. Because of the initial perturbations to the test particles, single armed phase spirals are very clear to see in the density plots $\rho(z, v_z)$. Although not as clear as those in the first row, single-armed phase spirals are also seen in $\langle v_R \rangle(z, v_z)$. We also show the same figure at $T\,=\,1.5\,\Gyr$, half a $\Gyr$ after the onset of the bar and spiral arms. Only the last panel of the density plot $\rho(z, v_z)$ shows spiral arm features now. The spiral structure in $\langle v_R \rangle(z, v_z)$ disappears. We can therefore draw two important conclusions. First, the one-armed spiral structure in this simulation lives longer in $\rho(z, v_z)$ than in $\langle v_R \rangle(z, v_z)$: the orbits in the outer disc can store this feature longer than the orbits in the inner disc, which can be attributed to the longer dynamical times in the outer disc. Second, the steadily rotating bar and spiral arm potentials used in this simulation do not produce new phase spiral features in the test particles, meaning that the joint action of an initial external vertical perturbation and internal plane-symmetric non-axisymmetries with steady pattern speeds do not create the observed vertical phase-space structures.
However, one has to keep in mind that the above conclusions only apply to the simplified specific simulation analyzed here : the external perturbation used was universal for the whole disc, which is not what is really expected from an intruder, and no additional self-consistent planar response perturbation was added as a response to the satellite. Moreover, no measurement uncertainties are considered on top of the simulation, which does not allow a direct visual comparison of, for instance, Figure~\ref{fig:phase_05gyr_pert} and Figure~\ref{fig:phase_vr}.

\begin{figure*}
	\includegraphics[scale=0.45]{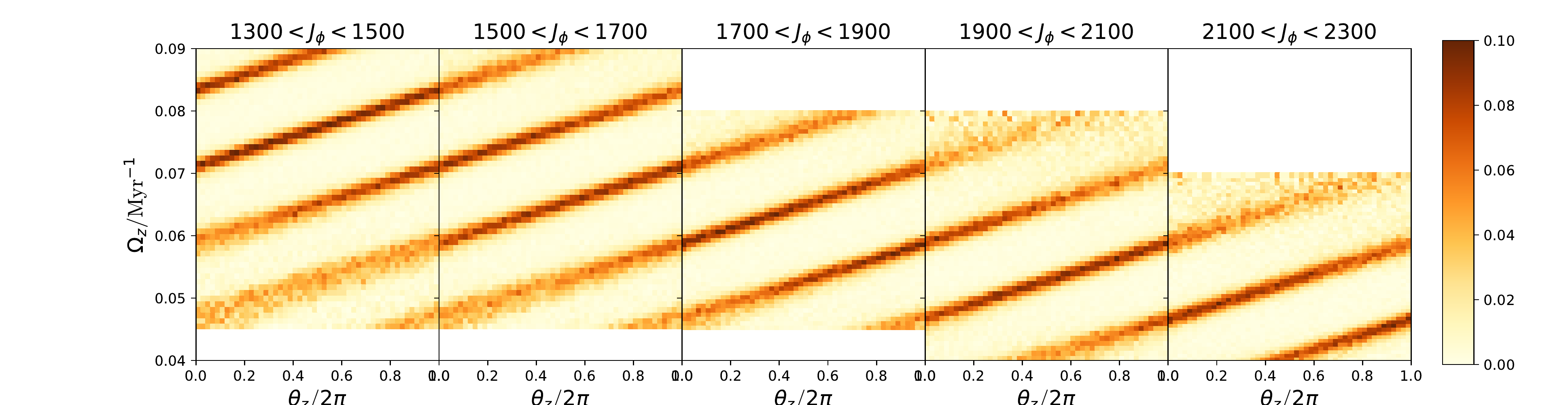}\\
         \vspace{0.1cm}
         \includegraphics[scale=0.45]{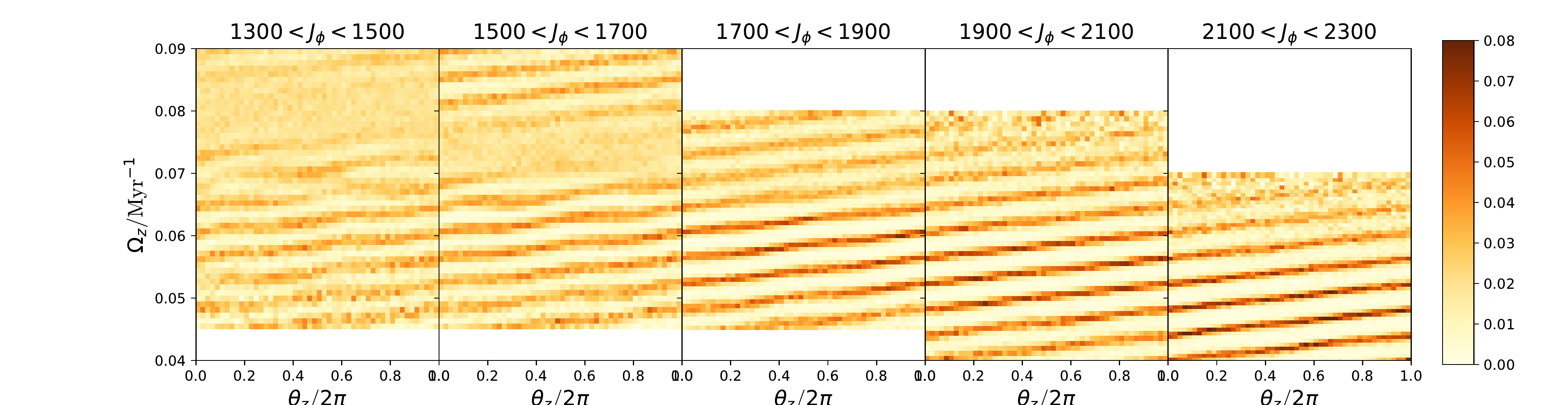}	
    \vspace{-0.2cm}
    \caption{Pseudo-star densities in the $\theta_z-\Omega_z$ plane for the model where an initial perturbation is added. Two time steps are showed, corresponding to $T\,=\,0.5\,\Gyr$ in the upper panel and $T\,=\,1.5\,\Gyr$ in the lower panel.  The re-normalization method is the same as in Figure~\ref{fig:thetazomegaz}. Clear stripes appear in the $\theta_z-\Omega_z$ space, and the number of the stripes increases at $T\,=\,1.5\,\Gyr$.}
    \label{fig:thetazomegaz_mock_05per}
    \vspace{-0.4cm}
\end{figure*}

We also plot the density of test particles with initial vertical perturbation in the $\theta_z-\Omega_z$ space at $T\,=\,0.5\,\Gyr$ in the upper panel of Figure~\ref{fig:thetazomegaz_mock_05per}. Clear stripes are shown in the $\theta_z-\Omega_z$ space. Since the perturbation is imposed to test particles together at $T\,=\,0.0\,\Gyr$, all the stripes show the same slopes which can be related to the age of this perturbation. The same plot is shown in the lower panel of Figure~\ref{fig:thetazomegaz_mock_05per}, but at $T\,=\,1.5\,\Gyr$. The number of stripes is increasing and they become thinner throughout the disc. The stripes, again in this plot, show the same slopes in all panels. The slopes become shallower than those at $T\,=\,0.5\,\Gyr$ because the perturbation occurred further away in the past.

\subsection{Time-varying strong bar toy-model}{\label{sec:str}}

At this stage, we have shown that internal plane-symmetric perturbations with constant pattern speeds, with or without an initial vertical velocity kick, are unable to produce a two-armed phase spiral as observed in the $\langle v_R \rangle(z, v_z)$ distribution of the inner disc in the Gaia data. Such a two-armed phase spiral is nevertheless a strong hint that a plane-symmetric perturbation is probably acting on disc stars in the Galaxy.

To demonstrate that such two-armed phase spirals can {\it in principle} be produced by plane-symmetric perturbations, we now explore a toy-model of a strong bar with a time-varying pattern speed, which should prevent the pseudo stars from finding an equilibrium within the potential of the strongest internal perturber since its pattern speed is constantly changing.

We therefore consider a bar with a large initial pattern speed and let the pattern speed decrease with time. The bar model we use in this toy-simulation is from \citet{Sormani2022}. The initial pattern speed is $\Omega_{\rm{b}}\,=\,-88\,\kms\,\kpc^{-1}$ at $T\,=\,0\,\Gyr$. Then it immediately starts to decrease and reach $\Omega_{\rm{b}}\,=\,-35\,\kms\,\kpc^{-1}$ at $T\,=\,6\,\Gyr$. Meanwhile, the mass and radial profile of the bar are set to be multiplied by factors 2.5 and 1.5 respectively, which roughly simulates the growth of the bar. The spiral arm model is also different from what we used in Section~\ref{sec:bar}. A four-armed spiral is adopted in this simulation, the scale length and scale height being $3.0\kpc$ and $0.5\kpc$ respectively, larger than those used in Section~\ref{sec:bar}. We then integrate test particle orbits for $\,8\,\Gyr$. This extreme model is not meant to represent the real Milky Way in any way, but is rather meant, as a proof of concept, to strongly disturb the disk in a plane-symmetric way whilst not letting the pseudo stars settling within the potential of the perturber whose pattern speed constantly varies with time.

\begin{figure*}
	\includegraphics[scale=0.50]{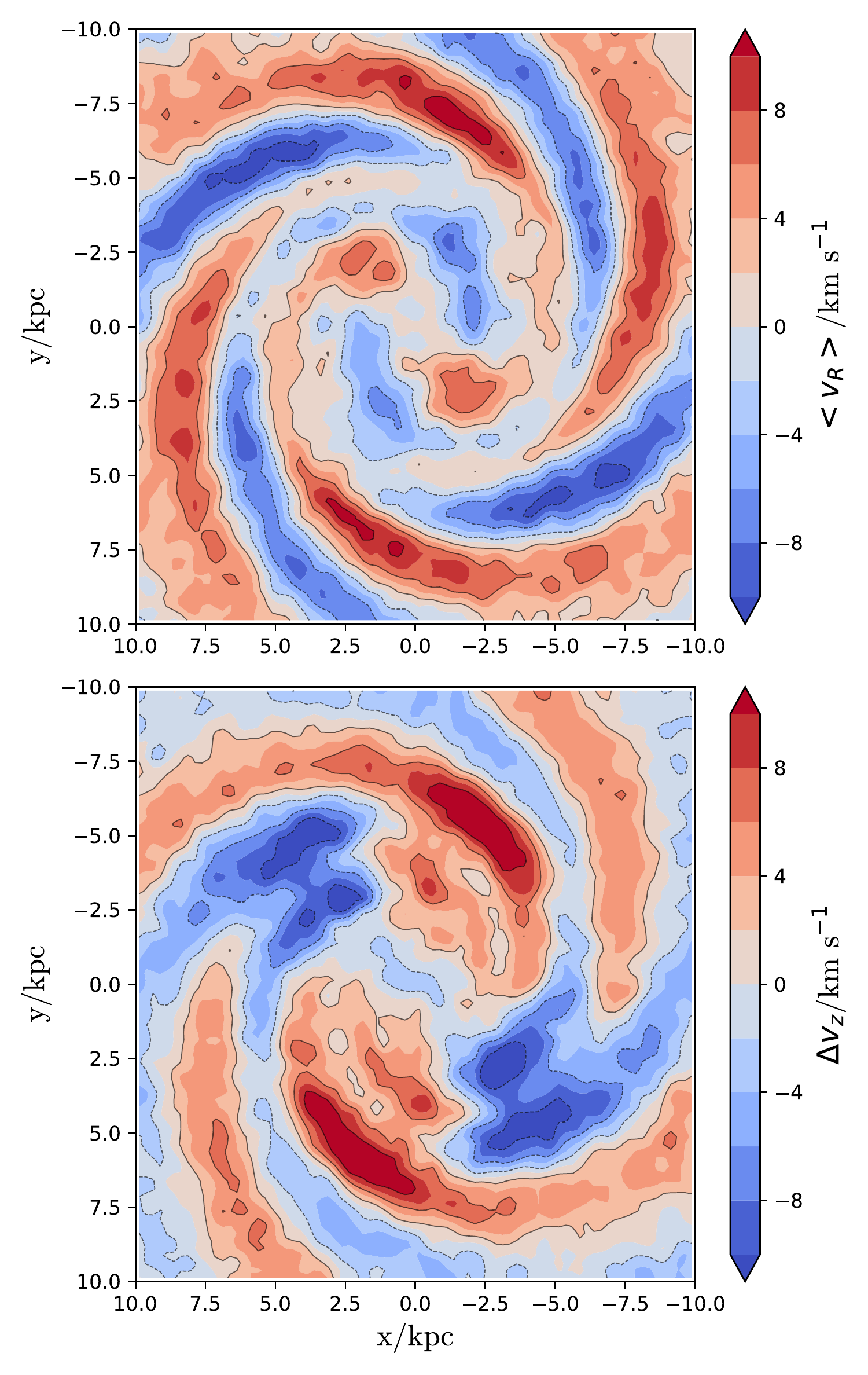}
	\hspace{-0.2cm}
        \includegraphics[scale=0.50]{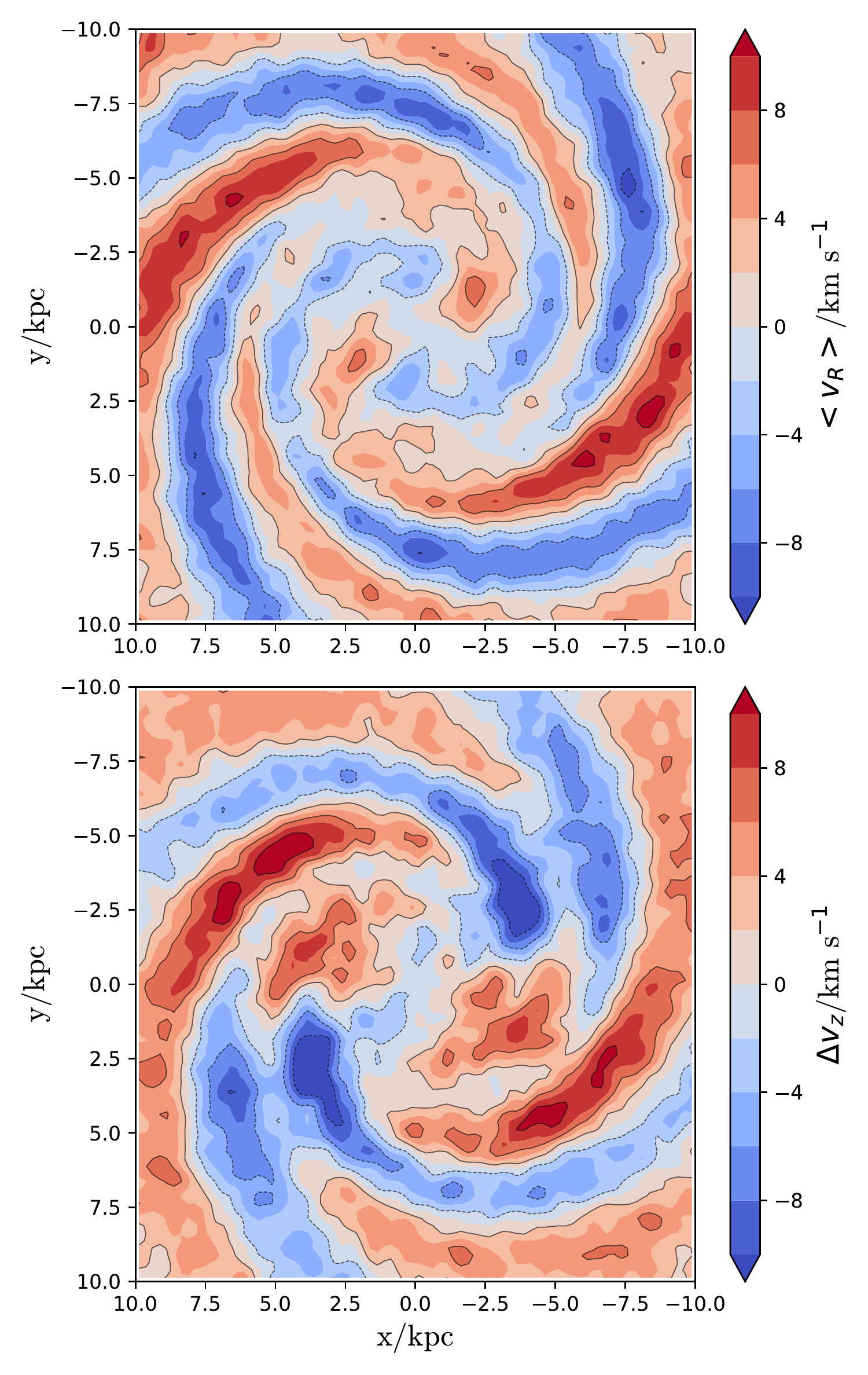}
    \vspace{-0.5cm}
    \caption{The face-on contour map of $\langle v_R \rangle$ and $\Delta v_z$ for the test particles of the slowing bar simulation at $\rm{T}\,=\,3.0\,\Gyr$ (bar pattern speed of $61.5\,\kms\,\kpc^{-1}$) in the left panel and $\rm{T}\,=\,3.5\,\Gyr$ (bar pattern speed of $57.1\,\kms\,\kpc^{-1}$) in the right panel respectively.  This map is also Gaussian smoothed with a filter width of $4\,\Delta\mbox{pix}$, with $\Delta \mbox{pix}=0.25\times0.25\,(\kpc\times\kpc)$.}
    \label{fig:contourf_dec}
    \vspace{-0.3cm}
\end{figure*}

In Figure~\ref{fig:contourf_dec}, the face-on contour map of $\langle v_r \rangle$ and $\Delta v_z$ for the test particles are shown at $\rm{T}\,=\,3.0\,\Gyr$ and $\rm{T}\,=\,3.5\,\Gyr$. Based on the simulation scheme, the pattern speed of the bar is approximately $\Omega_{\rm{b}}\,=\,-61.5\,\kms\,\kpc^{-1}$ at $\rm{T}\,=\,3.0\,\Gyr$ and $\Omega_{\rm{b}}\,=\,-57.1\,\kms\,\kpc^{-1}$ at $\rm{T}\,=\,3.5\,\Gyr$ respectively. We notice apparent quadruple features in $\Delta v_z$ between $R\,\sim\,3\,\kpc$ and the outer Lindblad resonance at $R\,\sim\,7\,\kpc$, which show that the strong bar plays an important role in the breathing mode of this model out to large radii. This breathing mode is additionally known to be non-linearly amplified in the joint presence of a strong bar and spiral arms \citep{monari2016b}.

\begin{figure*}
	\includegraphics[scale=0.45]{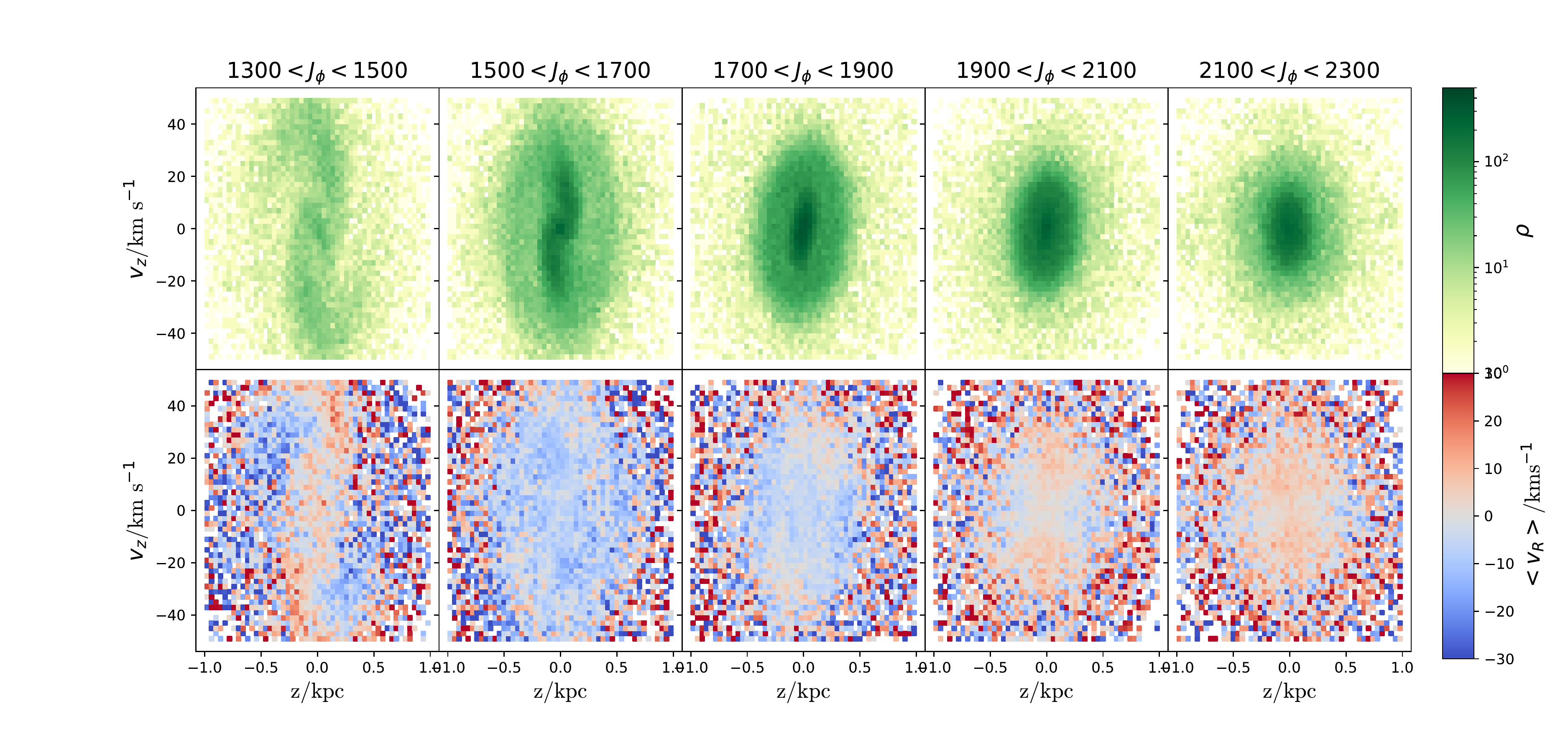}\\
        \vspace{-0.8cm}
        \includegraphics[scale=0.45]{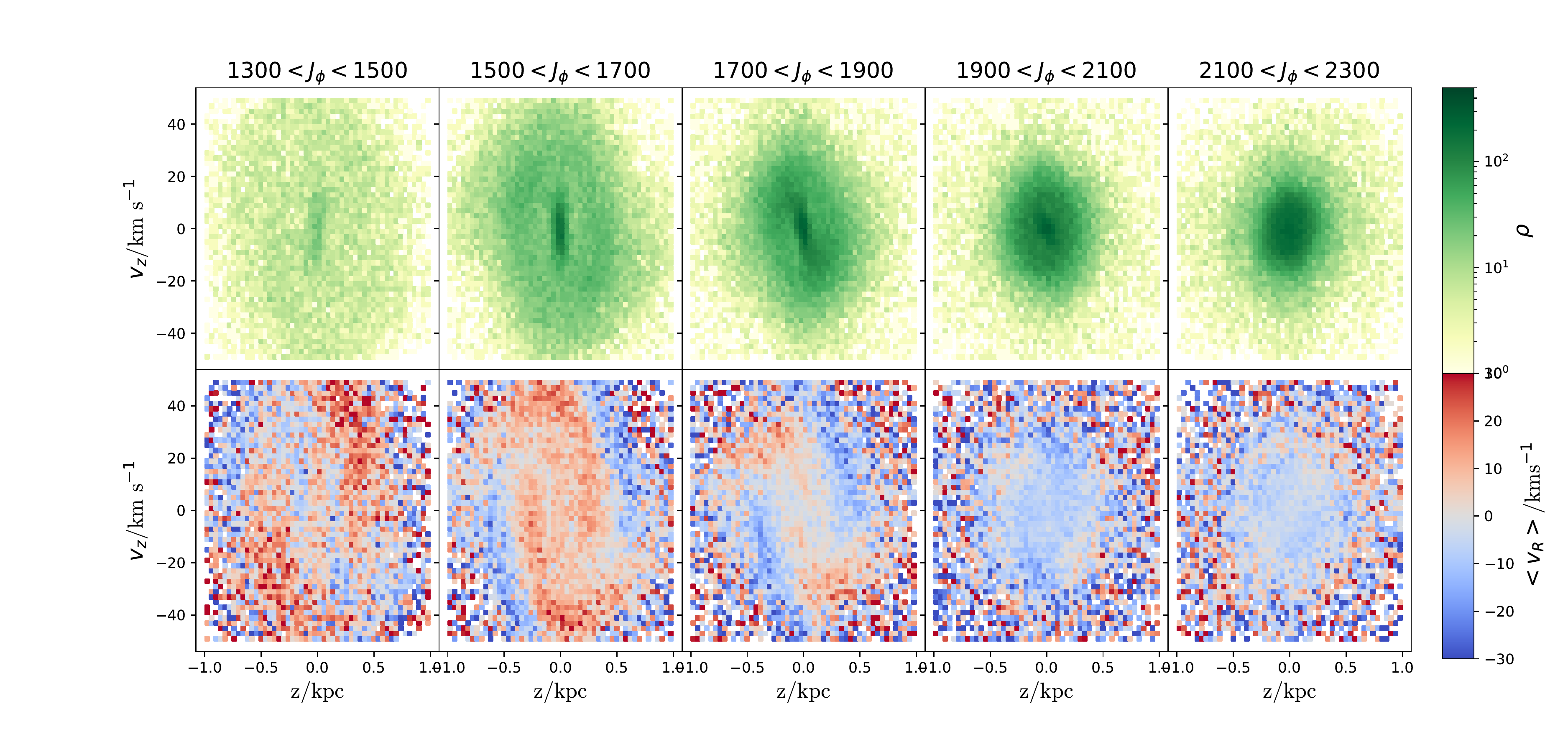}
	\vspace{-0.5cm}
    \caption{
    Maps of $\rho(z, v_z)$ and 
$\langle v_R \rangle(z, v_z)$ for test particles in the model with decreasing bar pattern speed, at $T\,=\,3.0\,\Gyr$ (upper grid) and $T\,=\,3.5\,\Gyr$ (lower grid). In each grid, the top row shows the density $\rho(z, v_z)$ and the bottom row $\langle v_R \rangle(z, v_z)$. The density plot $\rho(z, v_z)$ shows clear 2-armed spiral features in the first two panels at $T\,=\,3.0\,\Gyr$. The 2-armed spiral feature can also be seen in the first panel of $\langle v_R \rangle(z, v_z)$. At $T\,=\,3.5\,\Gyr$, the 2-armed spiral feature in $\rho(z, v_z)$ is eliminated but still clear in $\langle v_R \rangle(z, v_z)$.}
    \label{fig:phase_30gyr}
    \vspace{-0.3cm}
\end{figure*}

Figure~\ref{fig:phase_30gyr} then shows the $z-v_z$ phase space density of the pseudo stars at $T\,=\,3.0\,\Gyr$ in the upper panels and $T\,=\,3.5\,\Gyr$ in the lower panels. It shows clear 2-armed spiral features in the first two panels at $T\,=\,3.0\,\Gyr$. We also tested constant pattern speed bar models with the same spiral arm potential as in this model, and only this one produced the bisymmetric phase spiral, which is strong evidence that a bar with varying pattern speed is likely the culprit generating perturbations to the disc stars that lead to a 2-armed phase space spiral structure. The same pattern can also be seen in $\langle v_R \rangle(z, v_z)$ at $T\,=\,3.0\,\Gyr$ in the first panel. At $T\,=\,3.5\,\Gyr$, the phase space spirals disappear in $\rho(z, v_z)$, but are still clear to see in $\langle v_R \rangle(z, v_z)$, meaning that in this case, the two-armed phase spiral is sustained longer in $\langle v_R \rangle(z, v_z)$ than in density. 

Although this model is obviously not realistic at all as a model of the Milky Way, it still can tell us something useful. First of all, the decreasing of the bar's pattern speed appears to be capable of producing a 2-armed spiral feature. The same model without the change in pattern speed does not produce a phase spiral. So this result could indicate that the real Galactic bar has seen its pattern speed decrease from an initial large value up to its present day value. Second, the time scale to generate the phase spiral is long: in this simulation, the inner part of the Galaxy needs about 3 $\Gyr$ to generate the 2-armed phase spiral and the outer part needs a longer time of about $5-5.5$ $\Gyr$\footnote{Other phase space plots with different snapshots will be represented in the online supplementary files.}. Third, although the phase spiral patterns in $\rho(z, v_z)$ and $\langle v_R \rangle(z, v_z)$ have very strong relations, it is still possible to observe one without the other. The pattern in $\rho(z, v_z)$ disappears in less than $1\Gyr$ and the pattern in $\langle v_R \rangle(z, v_z)$ can actually survive for a longer time, which could explain why the observed 2-armed spiral features in Gaia data are clearer in $\langle v_R \rangle(z, v_z)$ than in $\rho(z, v_z)$, although it is premature to draw definitive conclusions based on the present toy-model. Finally, the features appear from inside-out, the outer part needing a longer time to generate this phase spiral feature. This can be associated to the decrease of the bar pattern speed, implying that the corotation and outer Lindblad resonance initially remain well within the inner disc before gradually moving out.

\section{Discussions and conclusions}{\label{sec:cons}}

In this work, we investigated the Gaia phase spiral structure based on observational data from Gaia DR3. We see a clear two-armed phase spiral structure in the inner disc (stars with low azimuthal action) in the $\langle v_R \rangle(z, v_z)$ map, which is the same as Figure 4 of \citet{Antoja2022}, and is a clear signature of a breathing wave around the disc. On the contrary, only a one-armed phase spiral structure is clearly seen in the $\rho(z, v_z)$ plots. The density in the $\Omega_z-\theta_z$ plane displays some straight lines with positive slopes in the outer parts of the Galaxy. In the inner Galaxy, the $\Omega_z - \theta_z$ density map presents overdensities in the form of a double arch, which is the consequence of a selection effect. 

The two-armed phase spiral observed in $\langle v_R \rangle(z, v_z)$  is a strong hint that the inner disk is influenced by an internal plane-symmetric perturbation. The obvious culprits for this are therefore the bar and spiral arms of the Galaxy.

In order to analyse the influence of such internal perturbations on the phase spirals observed with Gaia data, we used test particle simulations to investigate how such perturbations affect the vertical phase space structure of the Galactic disc. 

First, we showed that a realistic bar and spiral arm pattern with constant amplitude and pattern speed generates, as expected, non-zero mean $v_R$ values across the disc, while the mean $v_z$ remains null because the perturbations are plane-symmetric. The subtraction of the mean $v_z$ values in the North and South hemisphere above and below the Galactic plane demonstrate, as expected too, that the Galaxy undergoes a breathing mode, which is particularly visible in the inner Galaxy. However, this breathing mode only translates itself in an inclination of the $z-v_z$ ellipsoid but does not show any sign of a two-armed phase spiral.

We then added an initial perturbation to the vertical positions and velocities of the test particles to mimic the interaction of the disc with an external perturber. In that case, we found that a single armed phase spiral structure can be clearly seen at $T\,=\,0.5\,\Gyr$ for $\rho(z, v_z)$, 
$\langle v_R \rangle(z, v_z)$. On the contrary, at $T\,=\,1.5\,\Gyr$, there are no spiral features anymore in both $\langle v_R \rangle(z, v_z)$ and $\langle v_\phi \rangle(z, v_z)$ plot. Only the outermost azimuthal action bin of the density plot $\rho(z, v_z)$ shows a one-armed phase spiral feature. The number of stripes on the $\Omega_z-\theta_z$ plane also increases from $T\,=\,0.5\,\Gyr$ to $T\,=\,1.5\,\Gyr$, as expected from efficient phase-mixing.

We have thus shown that internal plane-symmetric perturbations with constant pattern speeds, with or without an initial vertical velocity kick, are unable to produce a two-armed phase spiral as observed in the $\langle v_R \rangle(z, v_z)$ distribution of the inner disc in the Gaia data. Such a two-armed phase spiral is nevertheless a strong hint that a plane-symmetric perturbation is probably acting on disc stars in the Galaxy. Therefore, we also tested the ability to disturb the disc with an (unrealistic) stronger bar with varying pattern speed, not allowing the pseudo stars to settle in the potential of the perturber due to a constantly evolving potential. The pattern speed of the bar decreases from a very high $\Omega_{\rm{b}}\,=\,-88\,\kms\,\kpc^{-1}$ at $T\,=\,0\,\Gyr$ to $\Omega_{\rm{b}}\,=\,-35\,\kms\,\kpc^{-1}$ at $T\,=\,6.0\,\Gyr$, whilst also increasing in mass and radial extent with factors of 2.5 and 1.5 respectively. Meanwhile, a four-armed spiral model is adopted in this simulation. In this toy-model, we find that the phase space density $\rho(z, v_z)$ does show clear 2-armed spiral features at $T\,=\,3.0\,\Gyr$,  which is strong evidence that a bar with a decreasing pattern speed is able to produce a 2-armed phase spiral. At $T\,=\,3.5\,\Gyr$, the 2-armed spiral in $\rho(z, v_z)$ disappears, but it is still clear in $\langle v_R \rangle(z, v_z)$, as in the Gaia observations of the inner disc. This result suggests that the pattern speed of the bar might not be constant over time in the real Galaxy. We note that it takes a long time to produce these two-armed phase spirals and that the phase spiral features appear from inside-out, probably related to the growth of the bar. The Milky Way's bar has been previously reported to both decrease in its pattern speed and grow in its radial amplitude \citep{Chiba2021a}, which is in line with our present findings.

However, the decreasing pattern speed model of the bar that we used in this work was a toy one, and to be taken only as a proof of concept. It by no means can trace the real evolution of the bar in the Galaxy. Given our present results, it is thus a natural perspective to next improve our approach and model the evolution history of the Milky Way bar in a more realistic model, to then compare the phase spiral patterns between the observations and the simulations. Moreover, the self-gravity, which is proposed to be crucial in shaping both the amplitude and pitch angle of the phase space spirals \citep{Widrow2023}, is neglected throughout this work. As it is reported by \citet{Widrow2023}, with the effect of self-gravity, the surface density increases rapidly and the disc experiences a compression in the $z-v_z$ phase space, which generates new vertical phase spirals during the evolution and leads to stronger phase spiral amplitude in contrast to the case without self-gravity. This also reveals two possible flaws in our simulation regarding the absence of self-gravity : one is that the signal of the phase spiral structure might be present but too weak to detect in the steady bar simulation due to the absence of self-gravity, although we suspect that phase spirals would be difficult to sustain once an equilibrium state is reached in terms of the bar's amplitude and pattern speed \footnote{Actually, \citet{Widrow2023} found stationary phase-spirals in some regions of the Galactic disc when treating the case of the self-gravitating response to a point mass travelling on a circular orbit in the disk, but this is different from a bar at equilibrium which is itself the perturber and the response while the point mass can be considered as an external perturber.}; the other is the fact that we only see the two-armed phase spiral once within specific $J_\phi$ bins in the time-varying bar model, which is probably related to the lack of self-gravity in our simulation, since the $N$-body simulations are also able to generate the phase spirals repeatedly throughout the evolution \citep{Hunt2022}. Thus, the self-gravity should be considered carefully in the future when the phase spiral features will be investigated quantitatively in comparison to the observational data.

We also note that the single-armed phase spiral pattern in the $z-v_z$ plane of the outer Galaxy is highly likely to be triggered by the impact of the Sagittarius (Sgr) dwarf galaxy on the Galactic disc \citep{Binney2018,Laporte2019,BlandHawthorn2021}. Despite our neglecting of self-gravity, our present findings suggest that a time-varying strong internal perturbation together with an external vertical perturbation such as the one from the Sgr dwarf could actually explain the transition between the two-armed and one-armed phase-spirals that is observed in Gaia data around the Solar radius. Hence, in future work, we plan to include a thorough and {\it realistic} analytic model of the bar including a decrease of its pattern speed and growth in the radial direction. Then, the Sgr dwarf galaxy and Large Magellanic Cloud potentials will also be added to the Galactic potential to investigate the total perturbation from both internal and external origins. Together with a theoretical endeavour to understand in detail the physics of the phase-space patterns found in the present work, as well as the effects of self-gravity which were currently neglected, this should allow us, in the end, to disentangle the origins of the subtle structure of the vertical phase space of our Galaxy, and infer from it its history from a holistic point of view, including both environmental and secular evolution effects.

\section*{Acknowledgements}
We thank the referees for their patient perusal and constructive advice, which have helped improving this manuscript. CL, AS, BF and GM acknowledge funding from the ANR grant N211483 MWdisc. SR acknowledges support from the Royal Society through the Newton International Fellowship NIF{\textbackslash}R1{\textbackslash}221850. CL thanks Eugene Vasiliev for advice about orbital integration with $AGAMA$. CL thanks Rimpei Chiba for the fruitful discussions about the bar model used in this work. This work presents results from the European Space Agency (ESA) space mission Gaia. Gaia data are being processed by the Gaia Data Processing and Analysis Consortium (DPAC). Funding for the DPAC is provided by national institutions, in particular the institutions participating in the Gaia MultiLateral Agreement (MLA). The Gaia mission website is https://www.cosmos.esa.int/gaia. The Gaia archive website is https://archives.esac.esa.int/gaia.
\section*{Data Availability}

The $AGAMA$ source code and phase space plots for pseudo stars at different snapshots are available in its online supplementary material. The observational sample and mock star catalogues will be shared on reasonable request to the corresponding author.



\bibliographystyle{mnras}
\bibliography{example} 




\appendix




\bsp	
\label{lastpage}
\end{document}